\documentclass[journal,draftcls,onecolumn,12pt,twoside]{IEEEtranTCOM}
\normalsize

\ifCLASSINFOpdf
\else
\fi

\usepackage{graphicx}
\usepackage{float}
\usepackage{amssymb}
\usepackage{amsmath}
\usepackage [autostyle, english = american]{csquotes}
\MakeOuterQuote{"}
\usepackage[bookmarks=false]{hyperref}
\usepackage{caption}
\begin{document}
%
\title
{ Multi-sequence Spreading Random Access \\[-0.5cm](MSRA) for Compressive Sensing-based \\[-0.5cm]Grant-free Communication}
%
%
%

\author{Ameha T.~Abebe,~\IEEEmembership{Member},
       and~Chung G.~Kang,~\IEEEmembership{Senior Member,~IEEE}
}

%
%
\markboth{IEEE Transactions on Communications}%
{Submitted paper}
%



\maketitle
\vspace*{-60pt}
\begin{abstract}
\vspace*{-10pt}
The performance of grant-free random access (GF-RA) is limited by the number of accessible random access resources (RRs) due to the absence of collision resolution. Compressive sensing (CS)-based RA schemes scale up the RRs at the expense of increased non-orthogonality among transmitted signals. This paper presents the design of multi-sequence spreading random access (MSRA) which employs multiple spreading sequences to spread the different symbols of a user as opposed to the conventional schemes in which a user employs the same spreading sequence for each symbol. We show that MSRA provides code diversity, enabling the multi-user detection (MUD) to be modeled into a well-conditioned multiple measurement vector (MMV) CS problem. The code diversity is quantified by the decrease in the average Babel mutual coherence among the spreading sequences. Moreover, we present a two-stage active user detection (AUD) scheme for both wideband and narrowband implementation. Our theoretical analysis shows that with MSRA activity misdetection falls exponentially while the size of GF-RA frame is increased. Finally, the simulation results show that about 82\% increase in utilization of RRs, i.e., more active users, is supported by MSRA than the conventional schemes while achieving the RA failure rate lower bound set by random access collision.
\end{abstract}


\begin{IEEEkeywords}
Compressive sensing; Grant-free random access; Machine-type communication; Multiple-sequence spreading random access (MSRA), Non-orthogonal multiple access (NOMA), Multiple measurement vector (MMV)
\end{IEEEkeywords}

\newpage 

%
\IEEEpeerreviewmaketitle

\section{Introduction}
\vspace*{-20pt}
%
%
%
%
\IEEEPARstart{G}{rant}-free random access  (GF-RA) allows immediate channel access as users transmit data, including control signaling in a single shot, without waiting for radio resource assignment [1-6]. In particular, for bursty and sporadic data transmission from massive machine type communication (mMTC) devices, GF-RA is dubbed as an efficient multiple access protocol. However, as contention resolution is not available in GF-RA, its performance is limited by access collisions, which in turn depend on the number of available $random$ $access$ $resources$ (RRs), defined herein as time, frequency, signatures (such as a spreading sequence, codebook, etc.), and combinations of these resources from which a user makes a selection for contention-based access [4-7]. 

By considering $intermittence$ in users activity [8], i.e., a very small fraction of the devices are active at GF-RA opportunity, sparse signal recovery, e.g., compressive sensing (CS)-based multiple access schemes, has been widely considered [9-16]. In the conventional CS-based multiple access, users spread their symbols, either in a time or frequency domain, by randomly selecting a sequence from a pool of non-orthogonal sequences [5, 6, 16, 17]. In order to quantify the important aspects of multiple access schemes in GF-RA, let  ${{N}_{R}}$ denote the number of orthogonal radio resources (e.g., resource elements in OFDMA systems) available for multiple access communication. On the basis of  ${{N}_{R}}$ orthogonal resources, ${{N}_{RR}}$  nonorthogonal RRs can be considered. For example, in spreading-based schemes, if the length of the spreading sequences and the number of available sequences are denoted as $M$ and ${{N}_{s}}$, respectively, then $N_R=M$ and $N_{RR}=N_s$. The CS-based multiple access allows a large radio resource $scaling$ $factor$ (SF) $K=\frac{N_{RR}}{{{N}_{R}}}$ as long as the $utilization$ $factor$ (UF) $L=\frac{N_{a}}{{{N}_{R}}}$ is below a certain threshold, where $N_a$ denotes the number of active users.  

Random access collision occurs when two or more UEs select the same RR. The collision rate, denoted as ${{P}_{c}}$, is defined as the ratio of the number of collided transmissions to the total number of simultaneous transmissions. Assuming that each user selects a random access resource with uniform and independent probability, if the number of active users follows a Poisson distribution with mean ${{\bar{N}}_{a}}$ [30], then the collision rate  is given as
\begin{equation}
{{P}_{c}}=1-{{e}^{-\frac{{{\bar{N}}_{a}}}{{{N}_{RR}}}}}. 
\end{equation}
From (1), it can easily be observed that the collision rate drops exponentially when the SF $K$ increases as ${{N}_{RR}}=K{{N}_{R}}$. Thus, more active users can be supported. However, the non-orthogonality (mutual correlation) among the user signals increases with SF resulting in increased packet detection error rate owing to multiple access interference (MAI) [7]. Therefore, the design goal of  a multiple access scheme which is designed for GF-RA is to increase the SF as well as UF whilst guaranteeing a successful recovery by mitigating the MAI. 

The most widely considered GF-RA  works like a slotted Aloha protocol [9-16]. Users then transmit multiple symbols in a synchronized manner whereby active/inactive users remain in the same state in a random access slot. For the slotted Aloha-based GF-RA, user activity can be learned (inferred) from the received multiple symbols that allow for the sparse signal recovery to be modeled into a multiple-measurement vector (MMV) class of CS problems [18-20]. 

In the case of conventional CS-based multiple access schemes [9-16], the same spreading sequence is used to spread each of symbols in GF data frame. We use the term $frame$ to refer to the block of data symbols transmitted by a user within a GF slot. The conventional schemes can then be referred to as a  single-sequence spreading random access (SSRA) schemes. In this paper, however, we consider a scheme that spreads the different symbols of a user in a GF slot with the different sequences. The situation wherein a user employs multiple sequences to spread symbols in a data frame can be referred to as multi-sequence spreading random access (MSRA) [21]. Note that the different sequences in MSRA spread  different symbols in a frame and thus, there is no loss in spectral efficiency as compared to SSRA. On the contrary, a user views different MAI for symbols spread by the different spreading sequences, i.e., MAI diversity is provided. In this paper, we show that this MAI diversity rendered by MSRA improves users' activity and data detection performance, ultimately increasing the GF access success rate. Conversely, we show that MSRA achieves a higher SF ($K$) and UF ($L$) than SSRA subject to the same random access success rate (packet error probability). The conference version of the ideas discussed in this paper are presented in [21] and [22]. While [21] introduced the idea of multiple sequence spreading, [22] evaluated its performance in multi-cell environment with comparison to other NOMA schemes. In particular, [22] showed the multi-cell interference doesn't alter the sparse structure in MSRA and can be modeled as a dispersed noise. This paper, in contrast with [21] and [22], presents a more general design of MSRA with narrow and wideband implementation considerations, theoretical performance analysis and a receiver structure that considers both channel estimation and MUD. In [27], a Zadoff-Chu (ZC) sequence-based preamble transmission and an SSRA-based data transmission are proposed. Therein, a two-stage active user detection (AUD) and channel estimation  (CE) is introduced. Even if [27] exploits data-aided AUD, the extrinsic activity information from SSRA-based data transmission for data-aided AUD is not as strong as compared to the proposed scheme. In this paper, we show that employment of the multiple sequences in the proposed scheme provides a stronger (more unique) data signature that allows for a reliable data-aided AUD while incurring a lower preamble overhead.     

In Section II, we review the basic concept of CS-based multiple access as a precursor to our new idea and discuss the motivation for our current proposal. In Section III, we present a baseline system model for the proposed RA scheme in contrast to the conventional CS-based multiple access scheme. Section IV presents a CS-based receiver structure, while Section V discusses a theoretical performance analysis. The simulation results are presented and discussed in Section VI. Finally, in the last section, a conclusion is drawn, and the direction of future work is suggested.

$Notation$: All boldfaced lowercase letters, e.g., $\mathbf{x}$, are vectors, and boldfaced uppercase letters, such as $\mathbf{A}$, are matrices. Italicized letters, e.g., $K$ and $x$, represent variables, while sets are denoted by calligraphic letters, such as $\mathcal{K}$. If $\mathcal{K}\subseteq \{1,2,\cdots ,n\}$ is a set of indices, then $\mathcal{\bar{K}}$ is its complement, i.e. $\mathcal{\bar{K}}\subseteq \{1,2,\cdots ,n\}\backslash \mathcal{K}$. Furthermore, $\Gamma (\cdot )$ denotes an indexing operator. Hence, ${{\mathbf{A}}_{\Gamma (\mathcal{K})}}$ is a submatrix built up by the columns of $\mathbf{A}$ indexed by a set $\mathcal{K}$. Similarly, ${{\mathbf{x}}_{\Gamma (\mathcal{K})}}$ is a subvector which contains elements of $\mathbf{x}$ specified by the indices in $\mathcal{K}$. For example, if $\mathcal{K}=\{1,3\}$, then ${{\mathbf{A}}_{\Gamma (\mathcal{K})}}$ and ${{\mathbf{x}}_{\Gamma (\mathcal{K})}}$ contain the first and third columns of $\mathbf{A}$, and elements of $\mathbf{x}$, respectively. To reduce notational crowding, the indexing operator can be dropped when this does not give rise to confusion, i.e., ${{\mathbf{A}}_{\mathcal{K}}}={{\mathbf{A}}_{\Gamma (\mathcal{K})}}$ and ${{\mathbf{x}}_{\mathcal{K}}}={{\mathbf{x}}_{\Gamma (\mathcal{K})}}$. Moreover, ${{\mathbf{0}}_{M\times 1}}$ and ${{\mathbf{1}}_{M\times 1}}$, are zeros and ones column vectors, respectively, with a dimension ($M\times 1$). Finally, ${{\mathbf{A}}^{\dagger }}$ is the Moore-Penrose pseudo-inverse of $\mathbf{A}$, and ${{\mathbf{A}}^{H}}$ is its Hermitian matrix.
\vspace*{-10pt}
\section{Background and Motivation}
\vspace*{-10pt}
In order to determine the limitations of the existing schemes, and the ways they can be improved, we first consider a very simplified CS and  slotted Aloha-based random access [3, 6] transmission model with a frame consisting of  ${{N}_{c}}$ symbols. In the simplified model, active users randomly select the sequences from a predefined pool of spreading sequences $\mathcal{S}=\left\{ {{\mathbf{s}}_{1}},{{\mathbf{s}}_{2}},\cdots ,{{\mathbf{s}}_{{{N}_{s}}}} \right\}$ where the number of available sequences ${{N}_{s}}$ is much larger than the sequence length ($M$). Moreover, when a narrowband transmission, i.e., a flat channel response, is considered the signal received for spread and then superposed $i$-th symbol is modeled as 
\begin{equation}
{{\mathbf{y}}_{i}}=\mathbf{SH}{{\mathbf{d}}_{i}}+{{\boldsymbol{\omega }}_{i}}=\mathbf{S}{{\mathbf{x}}_{i}}+{{\boldsymbol{\omega }}_{i}},\text{ }i=1,2,\cdots ,{{N}_{c}},           
\end{equation}
where $\mathbf{S}=\left[{{\mathbf{s}}_{1}},{{\mathbf{s}}_{2}},\cdots ,{{\mathbf{s}}_{{{N}_{s}}}} \right]$, and $\mathbf{H}$ is a diagonal matrix with the $n$-th diagonal element holding the channel gain coefficient ${{h}_{n}}$ associated with the $n$-th spreading sequence (channel of a user that chose the $n$-th spreading sequence). Additionally, ${{\mathbf{d}}_{i}}$ is a $N_s\times1$ symbol vector with its $n$-th element holding the $i$-th symbol from a user that selected the $n$-the sequence, and ${{\boldsymbol{\omega }}_{i}}\in {{\mathbb{C}}^{M}}$ is the corresponding noise vector. Furthermore, a matrix $\mathbf{A}=\mathbf{SH}$ represents the combined effect of the channel and spreading, while ${{\mathbf{x}}_{i}}=\mathbf{H}{{\mathbf{d}}_{i}}$ is a vector with channel-modulated symbols as its elements. Note that (2) is a sprase signal measurement problem as the number of nonzero valued elements of $\mathbf{d}_i$ is much less than its dimension, i.e.,  ${{\left\| {{\mathbf{d}}_{i}} \right\|}_{0}}\ll {{N}_{s}}$ where ${{\left\| \centerdot  \right\|}_{0}}$ is an ${{\ell }_{0}}$-norm operator which counts the number of nonzero elements. 

Let us denote the set of indices for the spreading sequences selected by active users be denoted by $\Lambda $ then for the $\text{supp}(\centerdot )$  operator which returns the indices of nonzero elements, we have $\Lambda =\text{supp}({{\mathbf{x}}_{1}})=\text{supp}({{\mathbf{x}}_{2}})=\cdots =\text{supp}({{\mathbf{x}}_{{{N}_{c}}}})$. In CS literature, $\Lambda $ is referred as $support$ $set$. Then, (2) can be rewritten as  ${{\mathbf{y}}_{i}}={{\mathbf{A}}_{\Lambda }}{{\mathbf{d}}_{i,\Lambda }}+{{\boldsymbol{\omega }}_{i}}={{\mathbf{S}}_{\Lambda }}{{\mathbf{x}}_{i,\Lambda }}+{{\boldsymbol{\omega }}_{i}}$, where ${{\mathbf{A}}_{\Lambda }}={{\mathbf{A}}_{\Gamma (\Lambda )}}$, ${{\mathbf{S}}_{\Lambda }}={{\mathbf{S}}_{\Gamma (\Lambda )}}$, ${{\mathbf{d}}_{i,\Lambda }}={{\mathbf{d}}_{i,\Gamma (\Lambda )}}$ and ${{\mathbf{x}}_{i,\Lambda }}={{\mathbf{x}}_{i,\Gamma (\Lambda )}}$. As per this simplified model, the first transmission can be used for channel measurement upon allowing users to transmit a known symbol, possibly a unit symbol, i.e., ${{d}_{1,n}}=1$ for $n\in \Lambda $, hence serving as preamble transmission. Here, ${{\mathbf{x}}_{1}}$ will simply represent the channel components, i.e., ${{x}_{1,n}}={{h}_{n}}$. For this simple system model, the support $\Lambda $ can be recovered by applying one of the greedy MMV algorithms on the measurements $\left\{ {{\mathbf{y}}_{i}}=\mathbf{S}{{\mathbf{x}}_{i}}+{{\boldsymbol{\omega }}_{i}} \right\}_{i=1}^{{{N}_{c}}}$ [17, 25]. Then, the corresponding channel can also be estimated via the least-square method as ${{\mathbf{\hat{x}}}_{1,\Lambda }}={{\left( {{\mathbf{S}}_{\Lambda }} \right)}^{\dagger }}{{\mathbf{y}}_{1}}$. Finally, the data symbols can then be estimated as  $\left\{ {{{\mathbf{\hat{d}}}}_{i,\Lambda }}={{\left( {{{\mathbf{\hat{A}}}}_{\Lambda }} \right)}^{\dagger }}{{\mathbf{y}}_{i}} \right\}_{i=2}^{{{N}_{c}}}$ where ${{\mathbf{\hat{A}}}_{\Lambda }}=diag({{\mathbf{\hat{x}}}_{1,\Lambda }}){{\mathbf{S}}_{\Lambda }}$ . Such support detection from data detection is generally referred as $data$-$aided$ $activity$ $detection$ [24] or $joint$ $activity$ $and$ $data$ $detection$ [23, 27]. 

The system model in (2) presents two limitations. First, it only applies to a narrowband communication system, such as NB-IoT [26], where the users' channel consists of a single tap. In a wideband systems with frequency selective channel wherein critical mMTC and ultra-reliable \& low latency communication (uRLLC) services are envisioned to be rendered, however, the columns of $\mathbf{A}$ are the spreading sequences convolved by the corresponding multi-tap channel vectors (time domain). In other words, the support information cannot be learned directly from data symbols transmission $\left\{ {{\mathbf{y}}_{i}}=\mathbf{A}{{\mathbf{d}}_{i}}+{{\boldsymbol{\omega }}_{i}} \right\}_{i=2}^{{{N}_{c}}}$, without knowing the channel.

   Second, the same measurement matrix $\mathbf{A}$ is employed for all data symbols transmission, i.e., $\left\{ {{\mathbf{y}}_{i}}=\mathbf{A}{{\mathbf{d}}_{i}}+{{\boldsymbol{\omega }}_{i}} \right\}_{i=2}^{{{N}_{c}}}$, which does not exploit code diversity. We show that by employing the different spreading sequences to spread different symbols, the measurement matrix $\mathbf{A}$ can now be made to vary among the different symbol measurements, i.e., $\left\{ {{\mathbf{y}}_{i}}={{\mathbf{A}}_{i}}{{\mathbf{d}}_{i}}+{{\boldsymbol{\omega }}_{i}} \right\}$. Given a set $\Lambda $, the 2-Babel mutual coherence of $\mathbf{A }$,  denoted as ${{\mu }_{2}}(\Lambda )$, is defined as the maximum correlation of an atom outside $\Lambda $ with atoms in $\Lambda $, which is expressed as follows: 
\begin{equation}
{{\mu }_{2}}(\Lambda )=\underset{k\notin \Lambda }{\mathop{\max }}\,\sqrt{\sum\limits_{i\in \Lambda }{{{\left| \left\langle {{\bf{a }}_{i}},{{\bf{a }}_{k}} \right\rangle  \right|}^{2}}}},
\end{equation}
where ${{\bf{a }}_{i}}$ is the $i$-th atom of $\mathbf{A }$ and $\left\langle {{\mathbf{a}}_{i}},{{\mathbf{a}}_{k}} \right\rangle =\frac{{{| \mathbf{a}_{i}^{H}{{\mathbf{a}}_{k}}|}}}{{{\left\| {{\mathbf{a}}_{i}} \right\|}_{2}}{{\left\| {{\mathbf{a}}_{k}} \right\|}_{2}}}$. In the case of spreading-based NOMA schemes, ${{\mu }_{2}}(\Lambda )$ can be interpreted as the worst non-orthogonality (the highest correlation) among the spreading sequences selected by active users and spreading sequences remained unselected. In this paper, we show that MSRA reduces the ${{\mu }_{2}}(\Lambda )$, i.e., averages out the non-orthogonality among transmissions. This phenomenon can be regarded as code diversity, which significantly increases the activity detection probability (performance of AUD) in compressive sensing signal recovery. Here, we want to note that the term $code$ $diversity$ is used to refer to the AUD performance improvement that has been achieved by MSRA.

    In this paper, we present a comprehensive system model and a GF transmission scheme that can be applied to both wideband and narrowband settings. In particular, a coarse support detection is performed that may include inactive users whilst insuring all active users are included, i.e., with a moderate false alarm and approximately null misdetection. From this coarse activity detection, the corresponding channel estimation is performed to form the estimated measurement matrices $\left\{ {{{\mathbf{\hat{A}}}}_{i}} \right\}$. Fine-tuning of the support detection, which is referred to as pruning in the CS literature, is then performed by considering the received multiple data symbols which are now modeled as $\left\{ {{\mathbf{y}}_{i}}={{{\mathbf{\hat{A}}}}_{i}}{{\mathbf{x}}_{i}}+{{\boldsymbol{\omega }}_{i}} \right\}$ in the receiver. Finally, channel estimation and data detection are performed based on the refined activity detection.

The contributions of this paper can be summarized as follows: i)	We present a system model for a slotted Aloha-based GF access scheme that can be generalized to both narrowband and wideband systems. The proposed transmission scheme allows for activity information to still be learned from data transmission even under multipath fading channel. ii) We demonstrate that upon multiple spreading sequences being employed by MSRA, a large radio resource scaling factor (SF) can be achieved while the MAI caused by non-orthogonal transmission is averaged out. Moreover, we parameterize the non-orthogonality between spreading signatures with the Babel mutual coherence [18]. In fact, it is shown that the Babel mutual coherence is reduced by MSRA, indicating an enhancement in code diversity. iii)	It is also shown that employment of the multiple spreading sequences as a signature fulfills the $innovation$ (variation) requirement for the multiple measurements [18, 19] so that the detection problem can be casted into a well-conditioned MMV problem. In addition, we present a theoretical analysis that the upper bound for the probability of activity misdetection decreases exponentially with the number of symbols in grant-free data frame as long as a constraint (which is less constraining for MSRA) on the size of support set is fulfilled.

\begin{figure}[h]
\centering
\vspace*{-15pt}
\includegraphics[width=4.5in,height=1.7in]{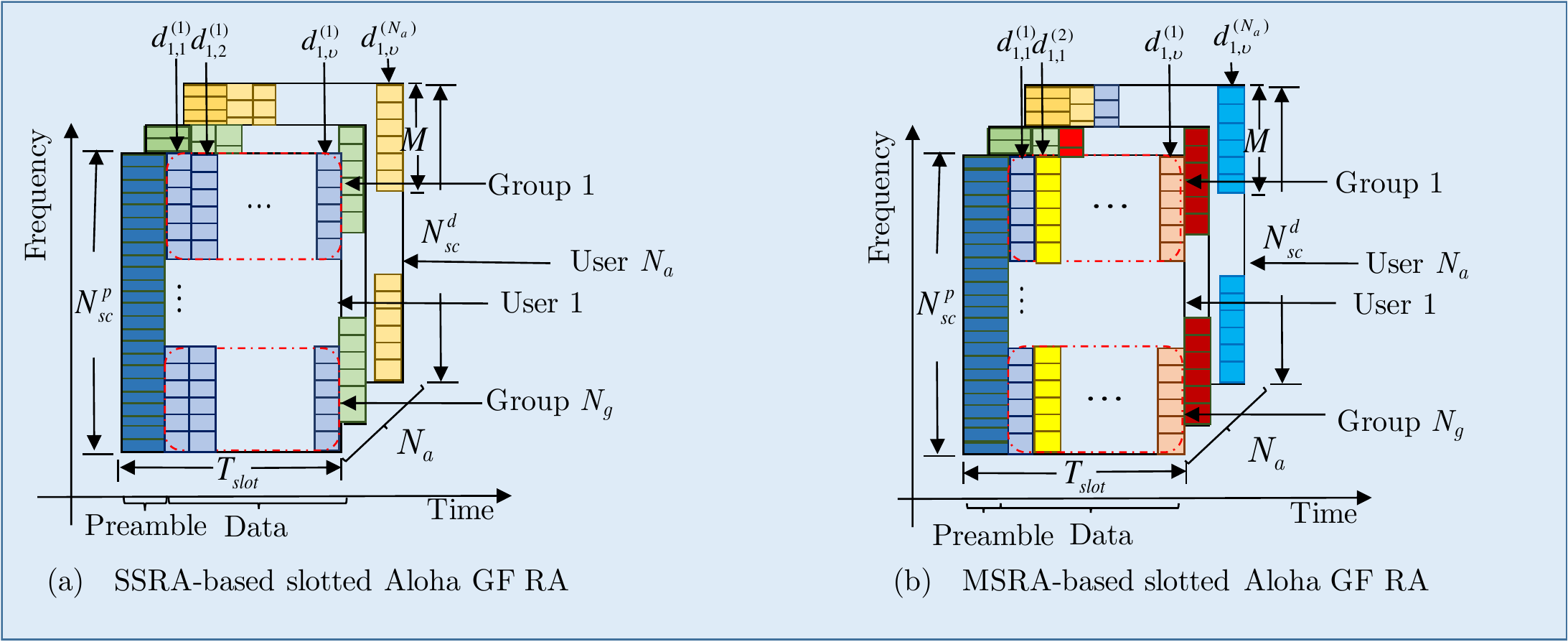}
\captionsetup{font=small,justification=centering}
\vspace*{-5pt}
\caption{\small{System model for preamble and data transmission: time-frequency random access resources}}
\vspace*{-40pt}
\label{fig_sim}
\end{figure}

\vspace*{-10pt}
\section{Comprehensive System Model for Multi-Sequence Random Access (MSRA)}
\vspace*{-10pt}
In this section, we discuss the design of multi-sequence spreading random access (MSRA) with a comprehensive system model that is general to be applied to both narrowband (NB) and wideband (WB) systems. In the sequel, we begin by discussing the preamble and data transmission model for the grant-free access which is shown in Fig. 1. Without loss of generality, we consider random access resources in OFDM time-frequency grid for preamble and data transmission slots that consist of $N_{sc}^{p}$ and $N_{sc}^{d}$ subcarriers, respectively. Furthermore, Fig. 1 depicts a grant-free slot of the multiple OFDM symbols for preamble and data transmission in time domain, respectively. We assume that the preamble and data symbols are transmitted within a channel coherence time    and hence, the channel estimated from the preamble transmission enables decoding the data symbols transmitted in the data transmission part of the GF slot. Furthermore, we present the system model in such a way that the conventional single-sequence spreading random access (SSRA)-based schemes [9-16],[23, 24] will be a special case of the proposed scheme. 

\subsection{Preamble Transmission}
\vspace*{-5pt}
An active user first selects a preamble sequences from a predefined pool of ${{N}_{p}}$  preambles, $\mathcal{P}=\left\{ {{\mathbf{p}}_{1}},{{\mathbf{p}}_{2}},\cdots ,{{\mathbf{p}}_{{{N}_{P}}}} \right\}$, where ${{\mathbf{p}}_{m}}\in {{\mathbb{C}}^{{{N}_{ZC}}\times 1}}$, $m=1,2,\cdots ,{{N}_{p}}$, is an ${{N}_{ZC}}$-length ZC sequence. Moreover, as $\mathcal{P}=\left\{ {{\mathbf{p}}_{1}},{{\mathbf{p}}_{2}},\cdots ,{{\mathbf{p}}_{{{N}_{P}}}} \right\}$ represents the time-domain preambles, the fast Fourier transformation (FFT)-transformed versions are fed to the inverse-FFT (IFFT) block of the OFDM system. In general, ${{N}_{P}}\gg {{N}_{ZC}}$ and thus, the preamble sequences are non-orthogonal to each other. When the number of preamble subcarriers, $N_{sc}^{p}$, is slightly greater than ${{N}_{ZC}}$, the first ($N_{sc}^{p}-{{N}_{ZC}}$) elements can be copied to the remaining subcarriers before the IFFT block of an OFDM transmitter. The generation of non-orthogonal ZC sequences from multiple ZC roots can be referred to in [27]. Let ${{\mathbf{p}}^{(k)}}\in \mathcal{P}$ denote a preamble sequence selected by the $k$-th active user, which has a time-domain multipath channel vector with delay spread length $\tau $, denoted as ${{\mathbf{\bar{h}}}^{(k)}}\in {{\mathbb{C}}^{\tau \times 1}}$. Assuming that ${{N}_{a}}$ active users randomly select one of the preambles from ${{N}_{p}}$ preambles, a received signal of the superposed preambles at the base station in the time domain is given as 
\begin{equation}
{{\mathbf{y}}_{p}}=\sum\limits_{k=1}^{{{N}_{a}}}{{{\mathbf{p}}^{(k)}}\otimes {{{\mathbf{\bar{h}}}}^{(k)}}}+{{\boldsymbol{\omega }}_{p}}=\mathbf{P\bar{h}}+{{\boldsymbol{\omega }}_{p}},
\end{equation}
where $\otimes$ is a circular convolution operator and ${{\boldsymbol{\omega }}_{p}}\sim \mathbb{C}\mathbb{N}(0,{{\sigma }^{2}}{{\mathbf{I}}_{{{N}_{ZC}}}})$   is the ambient noise with power ${{\sigma }^{2}}$ and  ${{\mathbf{I}}_{{{N}_{_{ZC}}}}}$ is an ${{N}_{ZC}}\times {{N}_{ZC}}$  identity matrix.  Let $\mathbf{P}$ denote a preamble matrix which has ${{N}_{p}}$ blocks of columns associated with each pilot sequence and their ($\tau -1$) cyclically-shifted versions, i.e., $\mathbf{P}=\left[ {{\mathbf{P}}_{1}},{{\mathbf{P}}_{2}},\cdots ,{{\mathbf{P}}_{{{N}_{P}}}} \right]\in {{\mathbb{C}}^{{{N}_{ZC}}\times \tau {{N}_{p}}}}$, with (${{N}_{zc}}\times \tau $) circulant matrix ${{\mathbf{P}}_{m}}=\left[ {{\mathbf{p}}_{m,1}},{{\mathbf{p}}_{m,2}},\cdots ,{{\mathbf{p}}_{m,\tau }} \right]$ formed by setting its first column by ${{\mathbf{p}}_{m,1}}={{\mathbf{p}}_{m}}$ and its $t$-th column ${{\mathbf{p}}_{m,t}}$ by a circularly-rotated version of  ${{\mathbf{p}}_{m}}$ with  ($t-1$) elements, $t\in \{2,3,\cdots ,\tau \}$. Meanwhile, let $\mathbf{\bar{h}}\in {{\mathbb{C}}^{\tau {{N}_{p}}\times 1}}$ denote a vector of the channel gain coefficients associated with each pilot. Note that $\mathbf{\bar{h}}$ is block-sparse in a sense that the number of nonzero values in it is lower than its dimension, i.e., ${{\left\| {\mathbf{\bar{h}}} \right\|}_{0}}\le \tau {{N}_{a}}\ll \tau {{N}_{p}}$, and these nonzero values are located as a group of $\tau $. As the circular convolution operation in (4) can be written as ${{\mathbf{p}}_{m}}\otimes {{\mathbf{\bar{h}}}_{m}}={{\mathbf{P}}_{m}}{{\mathbf{\bar{h}}}_{m}}$, where ${{\mathbf{\bar{h}}}_{m}}={{\mathbf{\bar{h}}}^{(k)}}$ if the user $k$ selects the $m$-th preamble; otherwise, ${{\mathbf{\bar{h}}}_{m}}=\mathbf{0}\in {{\mathbb{C}}^{\tau \times 1}}$, the received preamble signal in (4) can then be represented as ${{\mathbf{y}}_{p}}=\mathbf{P\bar{h}}+{{\boldsymbol{\omega }}_{p}}$.

Let us consider a set of ${{N}_{s}}$ base spreading sequences, denoted as ${{\mathcal{B}}_{s}}=\left\{{{\mathbf{s}}_{1}},{{\mathbf{s}}_{2}},\cdots ,{{\mathbf{s}}_{{{N}_{s}}}} \right\}$, where ${{\mathbf{s}}_{n}}\in {{\mathbb{C}}^{M}}$, $n=1,2,\cdots ,{{N}_{s}}$, is a spreading sequence of length $M$. Now, let ${{\mathcal{P}}_{s}}=\left\{ {{\mathcal{S}}^{(1)}},{{\mathcal{S}}^{(2)}},\cdots ,{{\mathcal{S}}^{({{N}_{T}})}} \right\}$ denote a pool of predefined ${{N}_{T}}$ sequence sets, each of which is constructed by selecting   sequences from ${{\mathcal{B}}_{s}}=\left\{ {{\mathbf{s}}_{1}},{{\mathbf{s}}_{2}},\cdots ,{{\mathbf{s}}_{{{N}_{s}}}} \right\}$. A spreading-sequence set modulated by a UE channel is referred to as a $signature$ since it, along with the preamble, is used to identify each transmission later for active-user detection (AUD). In this paper, we interchangeably use the same term signature to refer to the spreading-sequence set itself. Let us denote the $m$-th signature by ${{\mathcal{S}}^{(m)}}=\left\{ {{\mathbf{s}}_{m,1}},{{\mathbf{s}}_{m,2}},\cdots ,{{\mathbf{s}}_{m,\upsilon }} \right\}$ for $m=1,2,\cdots ,{{N}_{T}}$ consisting $\upsilon$ sequences, where ${{\mathbf{s}}_{m,l}}\in {{\mathbb{C}}^{M}}$ is taken from ${{\mathcal{B}}_{s}}$, i.e., ${{\mathbf{s}}_{m,l}}\in {{\mathcal{B}}_{s}}$. The radio resource SF ($K$) can be increased from $K={{N}_{s}}/M$ to $K={{N}_{T}}/M$ as long as  ${{N}_{T}}\gg {{N}_{s}}$ to reduce the access collision rate. 

Furthermore, we consider the case where there is one-to-one association between a signature and a preamble, i.e., ${{N}_{T}}={{N}_{p}}$, implying that if a user transmits the $m$-th  preamble sequence, it will also employ the $m$-th set of spreading sequences (spreading-sequence set). 
Active UEs transmit ${{N}_{c}}$ symbols within a GF slot, where each symbol is taken from the symbol alphabet $\mathcal{A}$. $M$-length spreading sequences are mapped to $M$ data subcarriers. In particular, ${{N}_{c}}$ symbols from the $k$-th UE are first split into ${{N}_{g}}={}^{{{N}_{c}}}/{}_{\upsilon }$ symbol groups, each group holding $\upsilon $ symbols as ${{\mathbf{d}}^{(k)}}=\left[ d_{1,1}^{(k)},d_{1,2}^{(k)}\cdots ,d_{1,\upsilon }^{(k)}|,\cdots ,|d_{{{N}_{g}},1}^{(k)},d_{{{N}_{g}},2}^{(k)}\cdots ,d_{{{N}_{g}},\upsilon }^{(k)} \right]$. Then, each symbol group is spread by the signature selected by the UE. For example, for the $k$-th UE that has selected the $m$-th signature ${{\mathcal{S}}^{(m)}}$, its $l$-th symbol in a group $i$, $d_{i,l}^{(k)}$, is spread using the $l$-th spreading sequence ${{\mathbf{s}}_{m,l}}$, $l=1,2,\cdots ,\upsilon $ in ${{\mathcal{S}}^{(m)}}$, i.e., representing each group spread as  $\left[ \begin{matrix} d_{i,1}^{(k)}{{\mathbf{s}}_{m,1}}, & \cdots  & ,d_{i,\upsilon }^{(k)}{{\mathbf{s}}_{m,\upsilon }}  \\
\end{matrix} \right]$. 

To give a toy example for a possible realization of the system model, suppose the preamble and data subcarriers are set to have the same SC spacing of 15KHz. Moreover, let the number of data and preamble SCs be $N_{sc}^{d}=N_{sc}^{p}=128$, i.e., a total bandwidth of 1.92MHz. Furthermore, let the GF RR consist of a preamble slot and a data slot with 14 OFDM symbols. If the spreading sequence length is set to be $M=32$, then the number of symbol groups can be set to ${{N}_{g}}=\frac{N_{sc}^{d}}{M}=4$ with each group of $\upsilon =14$ symbols. For example, an extended vehicular A (EVA) channel model for LTE entails five channel taps, i.e., $\tau =5$, for such bandwidth. Moreover, ZC-based sequences of length ${{N}_{ZC}}=127$ fit in the 128 subcarriers. The length of the preamble sequence can be adjusted without changing the bandwidth by adjusting the SC spacing in the preamble slot. An increase in the preamble length increases the associated preamble overhead (the length of the preamble slot in time domain), supporting a larger cell size or a greater number of users. 

From Fig. 1, it can be verified that the symbols in a symbol group can be mapped to different OFDM symbols at the same subcarrier locations so that they share the same channel. It should also be noted that a grant-free slot should be within a channel coherence time, i.e., ${{T}_{slot}}<{{t}_{c}}$ . Now since the spread symbols are mapped to subcarriers in a frequency domain, let us consider the frequency domain of the users' channels. Given the time-domain channel of user $k$, denoted as ${{\mathbf{\bar{h}}}^{(k)}}\in {{\mathbb{C}}^{\tau \times 1}}$, the circulant time-domain  channel matrix is given as ${{\mathbf{\bar{H}}}_{k}}=circ\left( {{[{{({{{\mathbf{\bar{h}}}}^{(k)}})}^{T}},{{({{\mathbf{0}}_{(N_{sc}^{d}-\tau)\times 1}})}^{T}}]}^{T}} \right)$, where ${{\mathbf{0}}_{(N_{sc}^{d}-\tau)\times 1}}$ is a $(N_{sc}^{d}-\tau)\times 1$ zero vector. Note that ${{[{{({{\mathbf{\bar{h}}}^{(k)}})}^{T}},{{({{\mathbf{0}}_{(N_{sc}^{d}-\tau)\times 1}})}^{T}}]}^{T}}$ is a zero-padded channel vector that is set to the first column of ${{\mathbf{\bar{H}}}_{k}}$. As ${{\mathbf{\bar{H}}}_{k}}$ is a circulant matrix, it can be expressed as ${{\mathbf{\bar{H}}}_{k}}={{\mathbf{W}}^{-1}}diag({{\mathbf{h}}^{(k)}})\mathbf{W}$ where $\mathbf{W}$ is an $N_{sc}^{d}\times N_{sc}^{d}$ discrete Fourier transform (DFT) matrix with its element at the $m$-th row and $n$-th column is given as  ${{w}_{m,n}}=\exp \left( {}^{-j2\pi mn}/{}_{N_{sc}^{d}} \right)$. Therefore, ${{\mathbf{h}}^{(k)}}$ is a frequency-domain channel vector over the $N_{sc}^{d}$ subcarriers, which remains unchanged over the OFDM symbols in the GF slot.

Furthermore, let ${{\mathbf{y}}_{i,l}}$ denote the received vector that corresponds to $\{d_{i,l}^{(k)}\}^{N_a}_{k=1}$ from the superimposed active UEs after Fourier transform and cyclic prefix (CP) removing. If $\mathbf{h}_{i,l}^{(k)}$ represents a part of ${{\mathbf{h}}^{(k)}}$ for the subcarriers in which the $i$-th group of symbols, $\{d_{i,l}^{(k)}\}$, are mapped, then, ${{\mathbf{y}}_{i,l}}$ is given as
\begin{equation}
{{\mathbf{y}}_{i,l}}=\sum\limits_{k=1}^{{{N}_{a}}}{diag(\mathbf{h}_{i,l}^{(k)})}{{\mathbf{s}}_{k,l}}d_{i,l}^{(k)}+{{\boldsymbol{\omega }}_{i,l}},  i=1,2,\cdots ,{{N}_{g}},    l=1,2,\cdots ,\upsilon,
\end{equation}
where ${{\mathbf{s}}_{k,l}}$ is the $l$-th spreading sequence in the sequence set selected by user $k$, and ${{\boldsymbol{\omega }}_{i,l}}\in {{\mathbb{C}}^{M\times 1}}$ is a noise vector. If the data symbols in a symbol group, $\left\{ d_{i,l}^{(k)} \right\}_{l=1}^{\upsilon }$, are mapped to the different OFDM symbols, but at the same subcarriers as shown in Fig. 1, then the channel remains the same during  $\upsilon $ symbols in a symbol group. A channel matrix $\mathbf{H}_{i}^{(k)}\in {{\mathbb{C}}^{\upsilon M\times \upsilon M}}=diag({{\left[ {{(\mathbf{h}_{i}^{(k)})}^{T}},{{(\mathbf{h}_{i}^{(k)})}^{T}},\cdots ,{{(\mathbf{h}_{i}^{(k)})}^{T}} \right]}^{\text{T}}})$ is then defined as a diagonal matrix formed by concatenating $\upsilon $ channel vectors ${{\mathbf{h}}_{i}^{(k)}}={{\mathbf{h}}_{i,1}^{(k)}}=\cdots ={{\mathbf{h}}_{i,\upsilon }^{(k)}}$. 

Let ${{\mathbf{S}}^{(m)}}\in {{\mathbb{C}}^{\upsilon M\times \upsilon }}$ denote a block-diagonal spreading matrix with diagonal blocks constructed from the sequences in the $m$-th signature ${{\mathcal{S}}^{(m)}}$. Moreover, let ${{\mathbf{S}}_{k}}$ denote the multi-sequence matrix that the $k$-th UE has selected. For example, if UE $k$ chose the $m$-th signature, ${{\mathbf{S}}_{k}}={{\mathbf{S}}^{(m)}}$. Furthermore, let ${{\mathbf{y}}_{i}}=\left[ \mathbf{y}_{i,1}^{T},\mathbf{y}_{i,2}^{T},\cdots ,\mathbf{y}_{i,\upsilon }^{T} \right]$ be a concatenation of the $\upsilon $ spread and superposed symbols. Then, it is expressed as 
\begin{equation}
{{\mathbf{y}}_{i}}=\sum\limits_{k=1}^{{{N}_{a}}}{{{\mathbf{H}}_{i,k}}}{{\mathbf{S}}_{k}}\mathbf{d}_{i}^{(k)}+{{\boldsymbol{\omega }}_{i}}, \text{ } i=1,2,\cdots ,{{N}_{g}},
\end{equation}
where $\mathbf{d}_{i}^{(k)}={{\left[ d_{i,1}^{(k)},d_{i,2}^{(k)}\cdots ,d_{i,\upsilon }^{(k)} \right]}^{\text{T}}}$ is the $i$-th symbol group for user $k$ and ${{\boldsymbol{\omega }}_{i}}=[{{({{\boldsymbol{\omega }}_{i,1}})}^{T}}$ ${{({{\boldsymbol{\omega }}_{i,2}})}^{T}}\cdots {{({{\boldsymbol{\omega }}_{i,\upsilon }})}^{T}} ]$. Note here that ${{\mathbf{y}}_{i}}={{\mathbb{C}}^{\upsilon M\times 1}}$ is a concatenation of $\upsilon $ spread and superposed symbols. It can in fact be written as ${{\mathbf{y}}_{i}}=\left[ \mathbf{y}_{i,1}^{T},\mathbf{y}_{i,2}^{T},\cdots ,\mathbf{y}_{i,\upsilon }^{T} \right]$, where $\left\{ {{\mathbf{y}}_{i,l}} \right\}_{l=1}^{\upsilon }$ are given in (5).  
\begin{figure}[t]
\centering
\includegraphics[width=5in,height=1.7in]{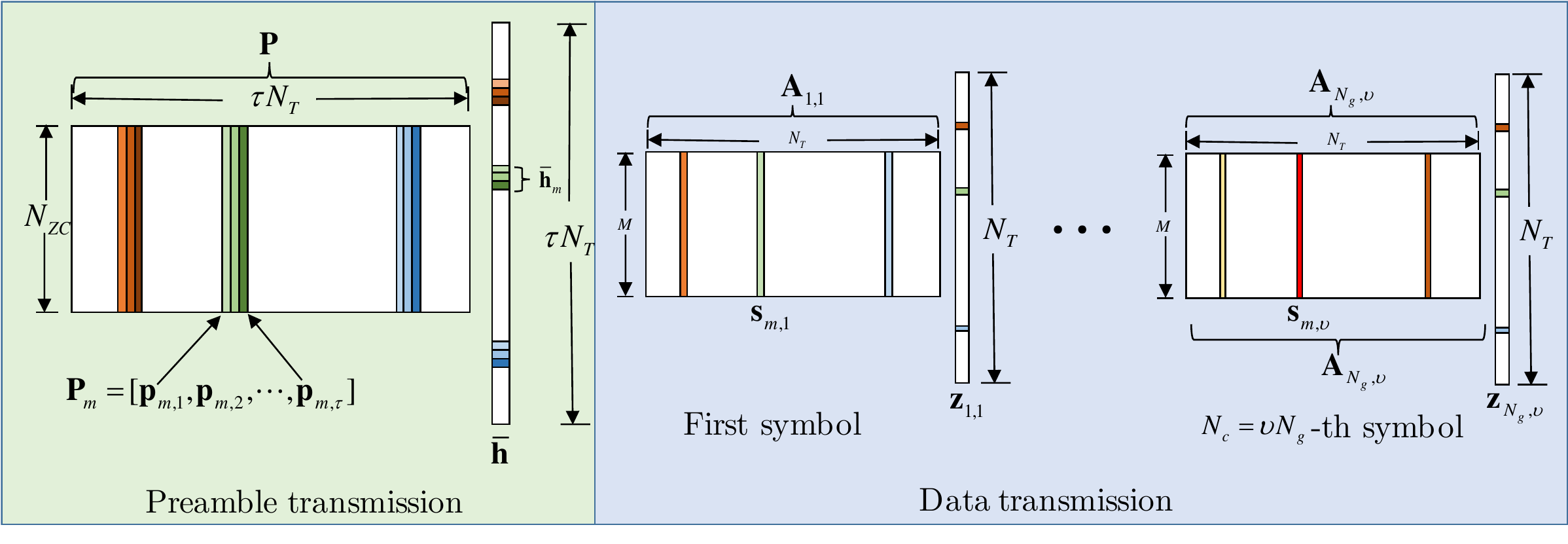}
\captionsetup{justification=centering}
\vspace*{-20pt}
\caption{\small{Illustration of multi-sequence spreading for the $i$-th symbol group in a data frame: $\tau $ = 3 and ${{N}_{a}}=3$}}
\vspace*{-30pt}
\label{fig_sim}
\end{figure}

To simplify notations, without loss of generality, let us consider the case where there is no collision during the selection of the spreading sequence sets. Furthermore, with a little abuse of notations, let $\mathbf{h}_{i}^{(m)}$ and $d_{i,l}^{(m)}$ be the frequency domain channel and data symbols, respectively, of users that have selected the $m$-th signature. If user $k$ selects the $m$-th signature, then  $\mathbf{h}_{i}^{(m)}=\mathbf{h}_{i}^{(k)}$ and $d_{i,l}^{(m)}=d_{i,l}^{(k)}$. Furthermore, for unselected signatures, $\mathbf{h}_{i}^{(m)}={{\mathbf{0}}_{M\times 1}}$ and $d_{i,l}^{(m)}=0$. Then, the received signal in (5) can be expressed in a matrix multiplication form as follows: 
\begin{equation}
{{\mathbf{y}}_{i,l}}={{\mathbf{A}}_{i,l}}{{\mathbf{z}}_{i,l}}+{{\boldsymbol{\omega }}_{i,l}}, \text{ }i=1,2,\cdots ,{{N}_{g}},\text{ }  l=1,2,\cdots ,\upsilon,
\end{equation}
where ${{\mathbf{A}}_{i,l}}\in {{\mathbb{C}}^{M\times {{N}_{T}}}}=\left[ diag(\mathbf{h}_{i}^{(1)}){{\mathbf{s}}_{1,l}},\cdots ,diag(\mathbf{h}_{i}^{({{N}_{T}})}){{\mathbf{s}}_{{{N}_{T}},l}} \right]$ and ${{\mathbf{z}}_{i,l}}\in {{\left\{ \mathcal{A},0 \right\}}^{{{N}_{T}}\times 1}}=$ $\text{}$ ${{\left[ d_{i,l}^{\text{(1)}}d_{i,l}^{(2)}\cdots d_{i,l}^{({{N}_{T}})} \right]}^{\text{T}}}$. It should be noted that since there exists an association of preambles and signatures, $\mathbf{\bar{h}}$ in (4)  and $\left\{ {{\mathbf{z}}_{i,l}} \right\}$ in (7)  share the same support, even if the former is a block-sparse signal. Fig. 2 illustrates the shared support for (4) and (7). Similarly, (6) can be rewritten in a matrix multiplication form as 
\begin{equation}
{{\mathbf{y}}_{i}}={{\mathbf{A}}_{i}}{{\mathbf{x}}_{i}}+{{\boldsymbol{\omega }}_{i}}, \text{ } i=1,2,\cdots ,{{N}_{g}},
\end{equation}
where ${{\mathbf{A}}_{i}}\in {{\mathbb{C}}^{\upsilon M\times \upsilon {{N}_{T}}}}=\left[ {{\mathbf{H}}_{i,1}}{{\mathbf{S}}^{(1)}},\cdots ,{{\mathbf{H}}_{i,{{N}_{T}}}}{{\mathbf{S}}^{({{N}_{T}})}} \right]$and ${{\mathbf{x}}_{i}}\in {{\left\{ \mathcal{A},0 \right\}}^{\upsilon {{N}_{T}}\times 1}}={{\left[ \mathbf{x}_{i,1}^{\text{T}},\cdots ,\mathbf{x}_{i,{{N}_{_{T}}}}^{\text{T}} \right]}^{\text{T}}}$ with $\mathbf{x}_{i,m}^{\text{T}}=\mathbf{d}_{i}^{(m)}$ holding the symbols in $i$-th symbols group from a user that has selected $m$-th signature. Note that (8) is a block-sparse signal measurement problem, since ${{\mathbf{x}}_{i}}\in {{\mathbb{C}}^{\upsilon {{N}_{T}}\times 1}}$ is a block-sparse vector with block size $\upsilon $ and ${{\left\| {{\mathbf{x}}_{i}} \right\|}_{0}}\le \upsilon {{N}_{a}}\ll \upsilon {{N}_{T}}$. Moreover, (8) is an MMV problem as $\text{supp}({{\mathbf{x}}_{i}})=\Lambda ,\text{ }\forall i\text{,}$ for a given support set $\Lambda $ that holds the indices for the block of elements. 
Now, single-sequence spreading-based random access (SSRA) which is discussed in [9-16], [23, 24, 27] can be considered as a special case of MSRA where the same sequence is used to spread each symbol in a symbol group. Fig. 3 illustrates the manner in which a symbol group could be spread in what we refer as a conventional single-sequence spreading random access (SSRA) (Fig. 3(a)) versus the multi-sequence spreading random access (MSRA) (Fig. 3(b)). The figure illustrates a symbol group of $\upsilon =4$, that is spread by SSRA and MSRA. In SSRA, all four symbols are spread by the same spreading sequence, while in MSRA they are spread by a signature of four different spreading sequences.

\section{Compressive Sensing-based Receiver for Grant-Free Random Access}
\vspace*{-5pt}
In this section, we discuss a receiver structure that exploits the various features of the GF-RA transmission discussed in the previous section. In the GF-RA, a receiver in the base station has to perform active user detection (AUD), the corresponding channel estimation (CE), and data symbols detection, all in one shot. This aims to exploit the structure in sparsity from preamble and data transmission in (4) and (8), enabling the AUD process that relies not only on the received preamble, but also on the received data symbols. As the users' channel is not yet known at the receiver end, however, the measurement matrices for symbol group transmissions, i.e., $\{{{\mathbf{A}}_{i}}\}_{i=1}^{{{N}_{g}}}$,  are also not known prior to AUD.

A widely employed approach to resolve this "$chicken$ $and$ $egg$" problem is a two-stage AUD [27], in which the first stage performs a crude estimation of active preambles from the received preamble signal while the second stage involves refining of these crude activity detection hypothesis by exploiting the sparsity structure in data symbols transmission. A block diagram of two-stage receiver is illustrated in Fig. 4. In particular, let $\mathcal{U}_{a}^{(1)}$ be a set of the indices for preamble sequences that are hypothesized as active in the first stage. The set $\mathcal{U}_{a}^{(1)}$ is first formed in such a way that there is very low misdetection while allowing a considerably large false alarm detection, i.e., $\left| \mathcal{U}_{a}^{(1)} \right|\gg {{N}_{a}}$. Using $\mathcal{U}_{a}^{(1)}$, then a channel of the hypothetically active users is estimated as $\left\{ \mathbf{\hat{h}}_{i}^{(m)} \right\}$ for $m\in {{\mathcal{U}}_{a}}$ and $i=1,2,\cdots ,\upsilon $. Furthermore, estimation of the measurement matrices $\left\{ {{{\mathbf{\hat{A}}}}_{i}} \right\}_{i=1}^{\upsilon }$ can now be formed from the estimated channel vectors $\left\{ \mathbf{\hat{h}}_{i}^{(m)} \right\}$. Finally, in the second stage, the AUD refines the set $\mathcal{U}_{a}^{(1)}$ by exploiting the transmission model in (8) while replacing the actual measurement matrices $\left\{ {{\mathbf{A}}_{i}} \right\}_{i=1}^{\upsilon }$ with their estimation, i.e., $\left\{ {{{\mathbf{\hat{A}}}}_{i}} \right\}_{i=1}^{\upsilon }$. As discussed in Section II, the receiver might not need to perform channel estimation before performing data-aided activity detection in a narrowband system where the channel is flat over the GF-RA frequency band. Therefore, a single-stage CE and AUD suffice. In the following, we discuss how the active user hypothesis set  $\mathcal{U}_{a}^{(1)}$ is formed from the received preamble transmissions and them how it s refined using data-aided AUD. 

\begin{figure}[t]
\centering
\includegraphics[width=4.5in,height=1.5in]{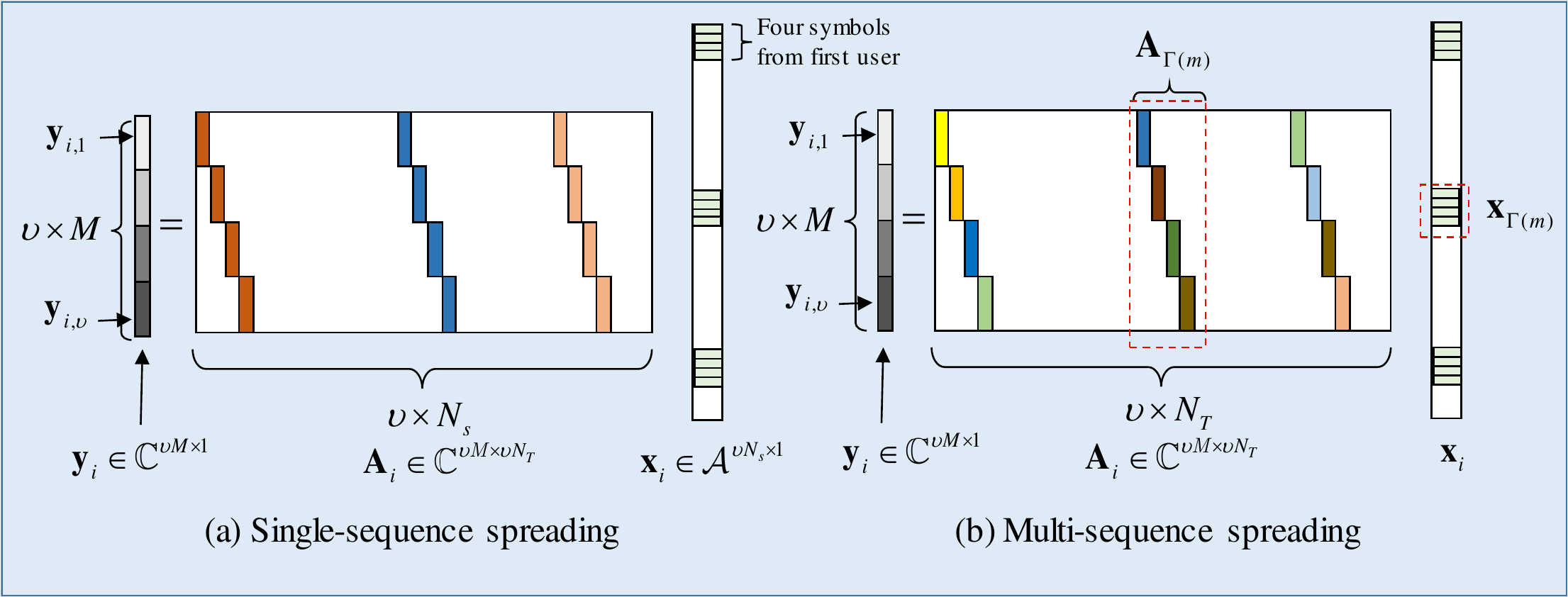}
\captionsetup{font=small,justification=centering}
\vspace*{-15pt}
\caption{\small{Random access with: single-sequence spreading vs. multi-sequence spreading: $\upsilon =4$ and ${{N}_{a}}=3$ }}
\vspace*{-30pt}
\label{fig_sim}
\end{figure}
 
\begin{figure}[h]
\centering
\includegraphics[width=4.5in,height=1in]{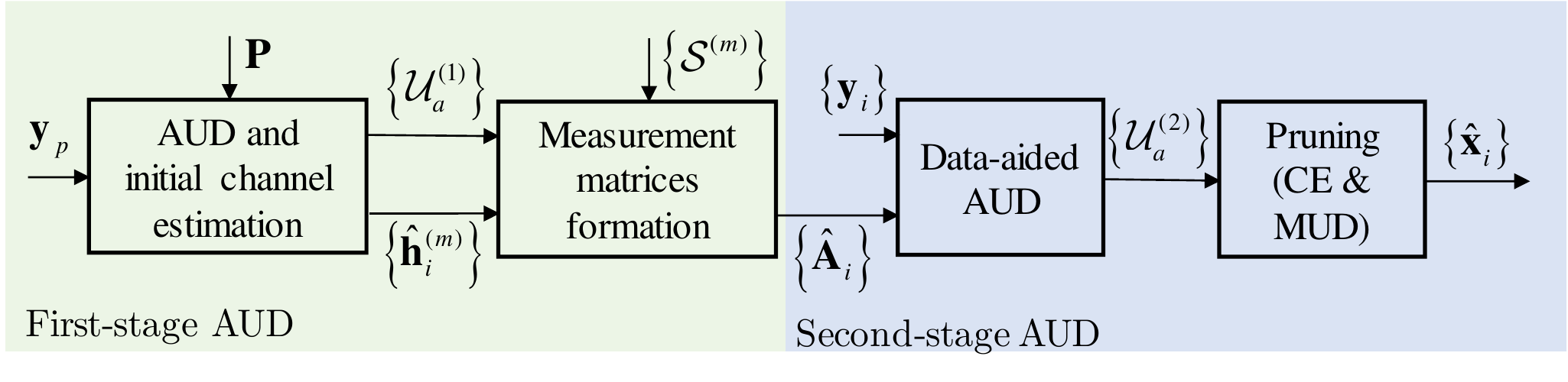}
\captionsetup{font=small,justification=centering}
\vspace*{-15pt}
\caption{\small{A two-stage receiver structure for grant-free random access scheme }}
\vspace*{-30pt}
\label{fig_sim}
\end{figure}

\subsection{Initial Active User Detection and Channel Estimation}
\vspace*{-5pt}
The initial AUD detection forms an activity hypothesis set $\mathcal{U}_{a}^{(1)}$ for the active users. For the sake of simplicity, we use the term $active$ $user$ to imply an active preamble that has been detected, ignoring the fact that an active preamble may mean the multiple active users under a collision scenario. As discussed in the previous section, the activity hypothesis set $\mathcal{U}_{a}^{(1)}$ has to include all the active users, i.e.,  $\Lambda \subseteq \mathcal{U}_{a}^{(1)}$, with a high probability. In this respect, we declare preambles that satisfy the following inequality with a threshold $\xi $: 
\begin{equation}
\left\|{{({{\mathbf{P}}_{m}})}^{H}}{{\mathbf{y}}_{p}} \right\|_{2}^{2}\ge \xi,
\end{equation}
as active, then the index $m$ is included in the hypothesis set $\mathcal{U}_{a}^{(1)}$. In [27], it is shown that setting the threshold $\xi$ low ensures all active users to be included in $\mathcal{U}_{a}^{(1)}$. In particular, setting $\xi $ below the expected power of channel $\left\| {{{\mathbf{\bar{h}}}}^{(k)}} \right\|_{2}^{2}$  and noise $\left\| {{\boldsymbol{\omega }}_{p}} \right\|_{2}^{2}$ leads to a low misdetection rate.  Once $\mathcal{U}_{a}^{(1)}$ is formed from (9), the least-square (LS) estimate of the time-domain channel vector  $\left( {\mathbf{\bar{h}}} \right)$ in (4) can be performed as ${{\mathbf{\tilde{h}}}_{\Gamma \left( \mathcal{U}_{a}^{(1)} \right)}}={{\left( {{\mathbf{P}}_{\Gamma \left( \mathcal{U}_{a}^{(1)} \right)}} \right)}^{\dagger }}{{\mathbf{y}}_{p}}$ and ${{\mathbf{\tilde{h}}}_{\Gamma \left( \mathcal{\bar{U}}_{a}^{(1)} \right)}}={{\mathbf{0}}_{\tau \left| \mathcal{\bar{U}}_{a}^{(1)} \right|\times 1}}$, where $\mathcal{\bar{U}}_{a}^{(1)}=\left\{ 1,2,\cdots ,{{N}_{T}} \right\}\backslash \mathcal{U}_{a}^{(1)}$ . It should be noted that for improved accuracy, a pruning operation, i.e., extraction of the highly likely correct support indices in   $\mathcal{U}_{a}^{(1)}$, can be identified by employing greedy CS algorithms such as OMP on the set $\mathcal{U}_{a}^{(1)}$ [27]. Then, the channel of these highly likely supports can be estimated separately. Moreover, the frequency-domain channel estimate, denoted as $\left\{ {{{\mathbf{\hat{h}}}}^{(m)}} \right\}$ for $m\in {{\mathcal{U}}_{a}}$, is computed with DFT as $\left\{ {{{\mathbf{\hat{h}}}}^{(m)}}=\mathbf{W}{{[{{\left( {{{\mathbf{\tilde{h}}}}^{(m)}} \right)}^{T}},{{\mathbf{0}}_{\left( N_{sc}^{d}-\tau  \right)\times 1}}]}^{T}} \right\}$ where ${{\mathbf{\tilde{h}}}^{(m)}}$ is a $\tau \times 1$ subvector of  ${{\mathbf{\tilde{h}}}^{(m)}}$ that belongs to the $m$-th index. Finally, indices of the estimated frequency-domain channel $\left\{ {{{\mathbf{\hat{h}}}}^{(m)}} \right\}$ are used to form the estimated measurement matrices $\left\{ {{{\mathbf{\hat{A}}}}_{i}} \right\}_{i=1}^{N_g }$ by following the matrix formation in (8). 
\vspace*{-25pt} 
\begin{figure}[t]
\centering
\includegraphics[width=3.8in,height=1.6in]{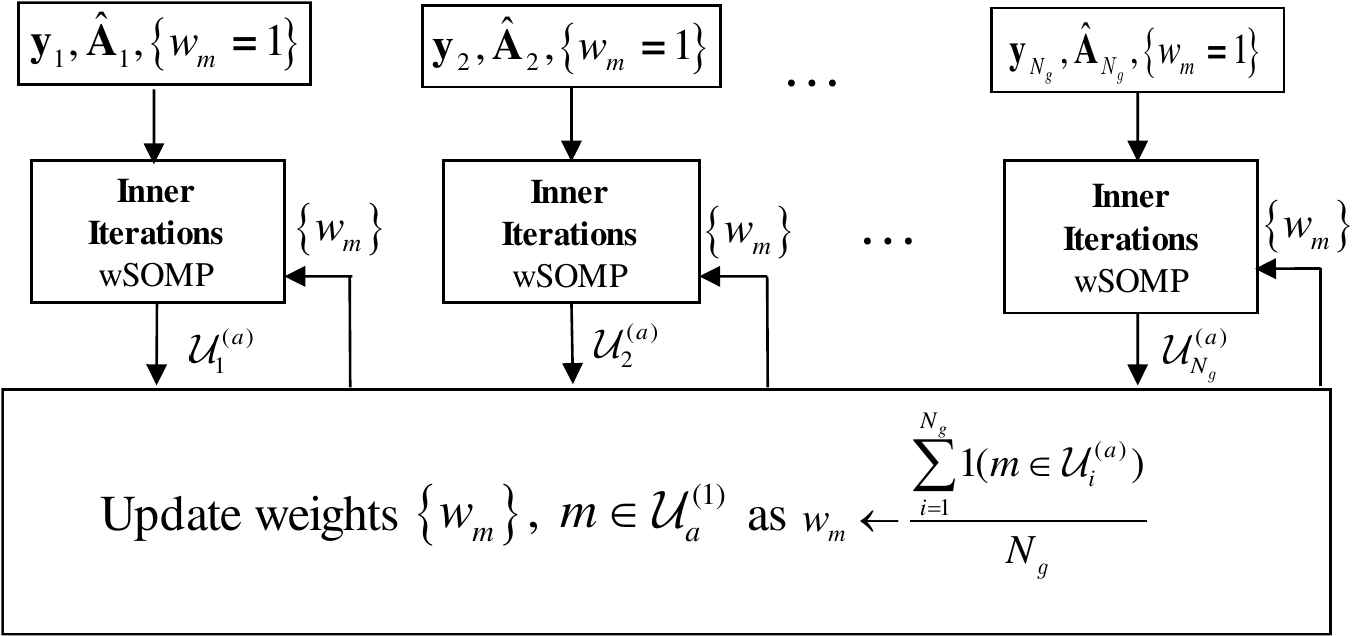}
\captionsetup{justification=centering}
\vspace*{-5pt}
\caption{\small{Block diagram of the IORLS-SOMP algorithm}}
\vspace*{-40pt}
\label{fig_sim}
\end{figure}

\subsection{Data-Aided Active User Detection}
\vspace*{-5pt}
Given the matrices $\left\{ {{{\mathbf{\hat{A}}}}_{i}} \right\}_{i=1}^{{{N}_{g}}}$ and the received vectors $\left\{ {{\mathbf{y}}_{i}} \right\}_{i=1}^{{{N}_{g}}}$, we aim at first recovering the support $\Lambda $ from the initial activity hypothesis $\mathcal{U}_{a}^{(1)}$. In this process, two additional pieces of information are used. First, we know that each vector ${{\mathbf{x}}_{i}}$ is block-sparse, i.e., for ${{\Lambda }_{i}}=\text{supp}({{\mathbf{x}}_{i}})$, we have $\left| {{\Lambda }_{i}} \right|\ll \upsilon {{N}_{T}}$. Furthermore, we have common support for measurements of each symbol group, i.e., $\Lambda ={{\Lambda }_{1}}=\cdots ={{\Lambda }_{{{N}_{g}}}}$. In this endeavor various MMV-based greedy algorithms [18-20] can be employed. For simplicity of the theoretical analysis, however, we consider a modified version of the iterative order-recursive least-square (IORLS) algorithm proposed in [20], as depicted by a block diagram in Fig. 5. Each iteration of the IORLS algorithm coupled with MSRA can be considered as ${{N}_{g}}$ operations of weighted simultaneous orthogonal matching pursuit (wSOMP),  each with $\upsilon $ measurements. 

The $weighted$-SOMP (wSOMP) refers to a modification of the SOMP algorithm in [18] where the atoms are selected via a weighted-correlation operations, rather than treating each atom similarly. In particular, we initialize the weights by ${{w}_{m}}=1$, $\forall m\in \mathcal{U}_{a}^{(1)}$ . Then wSOMP is performed over ${{N}_{g}}$ symbol groups in parallel. For the $i$-th symbol group, the residual is set to ${{\mathbf{r}}_{i}}={{\mathbf{y}}_{i}}$. Then, the index that maximizes the following is selected to be included in the set $\mathcal{U}_{i}^{(2)}$: 
\begin{equation}
m*=\arg \underset{m\in \mathcal{U}_{a}^{(1)}}{\mathop{\max }}\,\frac{\left\| {{w}_{m}}{{({{{\mathbf{\hat{A}}}}_{i,\Gamma (m)}})}^{H}}{{\mathbf{r}}_{i}} \right\|_{2}^{2}}{\left\| {{\left( {{{\mathbf{\hat{A}}}}_{i,\Gamma (m)}} \right)}} \right\|_{2}^{2}}.         
\end{equation}
Since (8) is a block-sparse signal measurement, ${{\mathbf{\hat{A}}}_{i,\Gamma (m)}}$ is block of columns of $\mathbf{\hat{A}}$ that corresponds with the $m$-th spreading signature, i.e., $\left\{ \upsilon (m-1)+1,\cdots ,\upsilon m \right\}$-th columns. This column block indexing is illustrated in Fig. 2. Therefore, (10) sums the square of the correlation of each column in ${{\mathbf{\hat{A}}}_{i,\Gamma (m)}}$ with the residual ${{\mathbf{r}}_{i}}$. Furthermore, the correlation is normalized by $\left\| {{\left( {{{\mathbf{\hat{A}}}}_{i,\Gamma (m)}} \right)}} \right\|_{2}^{2}$ in order to suppress the effect of channel estimation and the inaccuracy resulting thereafter while forming ${{\mathbf{\hat{A}}}_{i}}$. An LS-based estimation is then performed as ${{\mathbf{\hat{x}}}_{i,\Gamma \left( \mathcal{U}_{i}^{(2)} \right)}}={{\left( {{{\mathbf{\hat{A}}}}_{\Gamma \left( \mathcal{U}_{i}^{(2)} \right)}} \right)}^{\dagger }}{{\mathbf{y}}_{i}}$  and ${{\mathbf{\hat{x}}}_{i,\Gamma \left( \mathcal{\bar{U}}_{i}^{(2)} \right)}}={{\mathbf{0}}_{\left| \mathcal{\bar{U}}_{i}^{(2)} \right|\upsilon \times 1}}$. The residual ${{\mathbf{r}}_{i}}$ is then updated as ${{\mathbf{r}}_{i}}\leftarrow {{\mathbf{y}}_{i}}-{{\mathbf{\hat{A}}}_{i}}{{\mathbf{\hat{x}}}_{i}}$ upon projecting the received signal on the subspace that is orthogonal to the one spanned by columns of ${{\mathbf{\hat{A}}}_{\Gamma \left( \mathcal{U}_{i}^{(2)} \right)}}$. The wSOMP operation keeps on adding new index to support with (10) until $\left\| {{\mathbf{r}}_{i}} \right\|_{2}^{2}\le {{\xi }_{2}}$, where ${{\xi }_{2}}$ is a threshold that can be set to the noise variance. In particular, in [28] when a prior knowledge on the sparsity level is not available at a SOMP-based receiver, a stopping criteria for i.i.d Gaussian and arbitrary noise with variance ${{\sigma }^{2}}$ is computed as $\left\| {{\mathbf{r}}_{i}} \right\|_{2}^{2}\le \upsilon {{\sigma }^{2}}\left( M+2\sqrt{M\ln (M)} \right)$  and $\left\| {{\mathbf{r}}_{i}} \right\|_{2}^{2}\le \upsilon {{\sigma }^{2}}M$, respectively.  

After ${{N}_{g}}$ parallel wSOMP operations, the sets $\left\{ \mathcal{U}_{i}^{(2)} \right\}_{i=1}^{{{N}_{g}}}$ are now used to update the weights as 
\begin{equation}
{{w}_{m}}\leftarrow \frac{\sum\limits_{i=1}^{{{N}_{g}}}{\mathbf{1}(m\in \mathcal{U}_{i}^{(a)})}}{{{N}_{g}}},                   
\end{equation}
where $\mathbf{1}(.)$ denotes an indicator operator that returns 1 if its argument is true; otherwise it returns 0. Note that (11) counts the number of times the $m$-th signature is detected as active over ${{N}_{g}}$ symbol group measurements. In (11), since the weights encourage/suppress the signatures which are detected active/inactive in most of  the ${{N}_{g}}$ measurements, the sets $\left\{ \mathcal{U}_{i}^{(2)} \right\}_{i=1}^{{{N}_{g}}}$ converge to a certain set $\mathcal{U}_{a}^{(2)}$ [20]. A second round of CE is then performed on the converged activity set $\mathcal{U}_{a}^{(2)}$. It results in better performance as the received preamble signal is projected on to a subspace spanned by fewer and correct preambles (basis). Finally, the estimates of  ${{\mathbf{x}}_{i}}$ in (8), denoted as ${{\mathbf{\hat{x}}}_{i}}$, is computed as ${{\mathbf{\hat{x}}}_{i,\Gamma \left( \mathcal{U}_{a}^{(2)} \right)}}={{\left( {{{\mathbf{\hat{A}}}}_{\Gamma \left( \mathcal{U}_{a}^{(2)} \right)}} \right)}^{\dagger }}{{\mathbf{y}}_{i}}$ and ${{\mathbf{\hat{x}}}_{i,\Gamma \left( \mathcal{\bar{U}}_{a}^{(2)} \right)}}={{\mathbf{0}}_{\left| \mathcal{\bar{U}}_{a}^{(2)} \right|\upsilon \times 1}}$. 

 \section{Theoretical Performance Analysis}
 \vspace*{-5pt}
In this section, we provide a theoretical performance analysis of MSRA in terms of the correct support recovery rate, i.e., activity detection rate. In particular, we attempt to show the performance gain that MSRA provides, as compared to that of SSRA, owing to its inherent code diversity gain.  Accordingly, we consider the receiver structure discussed in the previous section for computing the upper bound on the probability of misdetecting at least one of the indices in the support, i.e., ${{P}_{mis}}=\Pr \{m\notin \mathcal{U}_{a}^{(2)}|m\in \Lambda \}$. 

First, let us focus on one of the ${{N}_{g}}$ wSOMP operations, and compute ${{P}_{mis}}$ by recognizing that wSOMP's initial operation, i.e., when ${{w}_{m}}=1,\text{ }\forall m$, is the same as SOMP [18] with $\upsilon $ measurements.  This is justified by how support indices are identified in (10). As it is discussed in the previous section, the use of multiple sequences in MSRA is equivalent to employing $\upsilon $ different measurement matrices, one for each compressed measurement. Let ${{\mathbf{\hat{a}}}_{(m-1)\upsilon +l,i}}$ denote the $((m-1)\upsilon +l)$-th atom (column) of  ${{\mathbf{\hat{A}}}_{i}}$. Therefore, ${{\mathbf{\hat{a}}}_{(m-1)\upsilon +l,i}}$ corresponds to the $l$-th sequence in the $m$-th signature. Then the measurement matrices are given by $\left\{ \mathbf{\Phi }_{i}^{(l)} \right\}_{l=1}^{\upsilon }$ where $\mathbf{\Phi }_{i}^{(l)}\triangleq \left[ {{{\mathbf{\hat{a}}}}_{l,i}},\cdots ,{{{\mathbf{\hat{a}}}}_{(m-1)\upsilon +l,i}},\cdots ,{{{\mathbf{\hat{a}}}}_{({{N}_{T}}-1)\upsilon +l,i}} \right]$. Note that $\mathbf{\Phi }_{i}^{(l)}$ corresponds to the estimated version of ${{\mathbf{A}}_{i,l}}$ in (6) except that it is formed from estimated channel. Therefore, (7) can be rewritten as the following measurements of $\upsilon $ symbols in the symbol group 
\begin{equation}
{{\mathbf{y}}_{i,l}}=\left( \mathbf{\Phi }_{i}^{(l)}-\mathbf{\Psi }_{i}^{(l)} \right){{\mathbf{z}}_{i,l}}+{{\boldsymbol{\omega }}_{i,l}}=\mathbf{\Phi }_{i}^{(l)}{{\mathbf{z}}_{i,l}}+{{{\boldsymbol{\bar{\omega }}}}_{i,l}},\text{ } l=1,2,\cdots ,\upsilon,  
\text{    }
\end{equation}
where $\mathbf{\Psi }_{i}^{(l)}=\mathbf{\Phi }_{i}^{(l)}-{{\mathbf{A}}_{i,l}}$ is a channel estimation error matrix and ${{\boldsymbol{\bar{\omega }}}_{i,l}}=\mathbf{\Psi }_{i}^{(l)}{{\mathbf{z}}_{i,l}}+{{\boldsymbol{\omega }}_{i,l}}$ is noise plus the estimation error vector. 

Let us analyze an arbitrary single symbol-group measurement so that the index '$i$' can be dropped from (10) and be rewritten as $\left\{{{\mathbf{y}}_{l}}={{\mathbf{\Phi }}^{(l)}}{{\mathbf{z}}_{l}}+{{{\boldsymbol{\bar{\omega }}}}_{l}} \right\}_{l=1}^{\upsilon}$. Furthermore, let $\mathsf{\ell}$ be an iteration counter for the inner wSOMP iterations. Assuming that the wSOMP operation accurately detects the signatures until the $\mathsf{\ell }$-th iteration, and a set $\mathcal{J}$ consists of signature indices ($m\in \mathcal{U}_{a}^{(1)}$) that are correctly identified at the $\mathsf{\ell }$-th iteration, i.e., $\mathcal{J}\subseteq \Lambda$ and $\left| \mathcal{J} \right|=\mathsf{\ell }$, the condition for the SOMP to fail to detect a correct atom at the $\text{(}\mathsf{\ell }+1)$-th iteration is explained in [18].  Using the same approach under the assumption that the different measurement matrices $\left\{ {{\mathbf{\Phi }}^{(l)}} \right\}_{l=1}^{\upsilon }$ are employed, the condition for misdetection (failure) in [18] is modified as 
\begin{equation}
\underset{m\in \Lambda \backslash \mathcal{J}}{\mathop{\min }}\,\sum\limits_{l=1}^{\upsilon }{{{\left| \left\langle \boldsymbol{\varphi }_{m}^{(l)},\mathbf{Q}_{\mathcal{J}}^{(l)}\boldsymbol{\varphi }_{m}^{(l)}{{z}_{l,m}} \right\rangle  \right|}^{2}}}-\underset{k\in \bar{\Lambda }}{\mathop{\max }}\,\sum\limits_{l=1}^{\upsilon }{{{\sum\limits_{m\in \Lambda \backslash \mathcal{J}}{\left| \left\langle \boldsymbol{\varphi }_{m}^{(l)},\mathbf{Q}_{\mathcal{J}}^{(l)}\boldsymbol{\varphi }_{k}^{(l)}{{z}_{l,k}} \right\rangle  \right|}}^{2}}}<\eta,
\end{equation}
where $\boldsymbol{\varphi }_{m}^{(l)}$ is the $m$-th atom in ${{\boldsymbol{\Phi }}^{(l)}}$ and $\mathbf{Q}_{\mathcal{J}}^{(l)}={{\mathbf{I}}_{M}}-\mathbf{P}_{\mathcal{J}}^{(l)}$  is a projection matrix to the subspace that is orthogonal to that spanned by the submatrix consisting of atoms in ${{\mathbf{\Phi }}^{(l)}}$ indexed by $\mathsf{\mathcal{J}}$ where $\mathbf{P}_{\mathcal{J}}^{(l)}=\mathbf{\Phi }_{\Gamma (\mathcal{J})}^{(l)}{{(\mathbf{\Phi }{{_{\Gamma (\mathcal{J})}^{(l)}}^{H}}\mathbf{\Phi }_{\Gamma (\mathcal{J})}^{(l)})}^{-1}}\mathbf{\Phi }{{_{\Gamma (\mathcal{J})}^{(l)}}^{H}}$. Furthermore, $\left\langle \mathbf{a},\mathbf{b} \right\rangle =\frac{{{| {{\mathbf{a}}^{H}}\mathbf{b} |}}}{{{\left\| \mathbf{a} \right\|}_{2}}{{\left\| \mathbf{b} \right\|}_{2}}}$ is a correlation between vectors $\mathbf{a}$ and $\mathbf{b}$. Note that the residual in the SOMP algorithm that corresponds to the $l$-th symbol detection is given as ${{\mathbf{r}}_{l}}=\mathbf{Q}_{\mathcal{J}}^{(l)}{{\mathbf{y}}_{l}}$. The first term in (13) is the minimum correlation between correct atoms (atoms in $\Lambda $) and the residual at the $\mathsf{\ell }$-th iteration, and ${{z}_{l,m}}$ denotes the $m$-th element of ${{\mathbf{z}}_{l}}$. Accordingly, the second term represents the maximum correlation between a wrong atom ($k\in \bar{\Lambda }$ where $\bar{\Lambda }=\left\{ \mathcal{U}_{a}^{(1)}\backslash \Lambda  \right\}$) and the residual. For the worst case, assuming that the wrong atoms destructively correlate with the remaining unidentified atoms, the second term is summed over all atoms in $\Lambda \backslash \mathcal{J}$. Furthermore, we set the worst correlation of the noise to the residual as $\eta =\sum\limits_{l=1}^{\upsilon }{\sum\limits_{m\in \Lambda }{{{\left| \left\langle \boldsymbol{\varphi }_{m}^{(l)},\mathbf{Q}_{\mathcal{J}}^{(l)}{{\boldsymbol{\varphi }}_{m}}{{{\bar{\omega }}}_{l,m}} \right\rangle  \right|}^{2}}}+}\sum\limits_{l=1}^{\upsilon }{\sum\limits_{i\in \bar{\Lambda }}{{{\left| \left\langle \boldsymbol{\varphi }_{m}^{(l)},\mathbf{Q}_{\mathcal{J}}^{(l)}\boldsymbol{\varphi }_{m}^{(l)}{{{\bar{\omega }}}_{l,m}} \right\rangle  \right|}^{2}}}}$. Assuming perfect power control with constant-modulus modulation alphabets, the transmitted symbols $\{{{z}_{l,m}}\}$ possess unit power and can therefore be omitted from the first term. Similarly, as noise multiplies with the atoms in $\bar{\Lambda }$, the second term can be maximized by considering the largest noise term, i.e., $\left| {{\omega }_{\max }} \right|=\underset{k\in \bar{\Lambda }}{\mathop{\max }}\,\left| \sum\limits_{l=1}^{\upsilon }{{{{\bar{\omega }}}_{l,k}}} \right|$ where ${{\bar{\omega }}_{l,k}}$ denotes the $k$-th element of ${{\boldsymbol{\bar{\omega }}}_{l}}$. Then, (13) can be modified to
\begin{equation}
\underset{m\in \Lambda \backslash \mathcal{J}}{\mathop{\min }}\,\sum\limits_{l=1}^{\upsilon }{{{\left| \left\langle \boldsymbol{\varphi }_{m}^{(l)},\mathbf{Q}_{\mathcal{J}}^{(l)}\boldsymbol{\varphi }_{m}^{(l)} \right\rangle  \right|}^{2}}}-{{\left| {{\omega }_{\max }} \right|}^{2}}\underset{k\in \bar{\Lambda }}{\mathop{\max }}\,\sum\limits_{l=1}^{\upsilon }{{{\sum\limits_{m\in \Lambda \backslash \mathcal{J}}{\left| \left\langle \boldsymbol{\varphi }_{m}^{(l)},\mathbf{Q}_{\mathcal{J}}^{(l)}\boldsymbol{\varphi }_{k}^{(l)} \right\rangle  \right|}}^{2}}}>\eta .
\end{equation}
The condition in (14) should be fulfilled until all atoms in the support are identified, i.e., $\mathcal{J}=\Lambda $. To compute the probability with which (14) is not satisfied, we define variables, ${{c}_{\Lambda }}\triangleq \underset{m\in \Lambda \backslash \mathcal{J}}{\mathop{\min }}\,\sum\limits_{l=1}^{\upsilon }{{{\left| \left\langle \boldsymbol{\varphi }_{m}^{(l)},\mathbf{Q}_{\mathcal{J}}^{(l)}\boldsymbol{\varphi }_{m}^{(l)} \right\rangle  \right|}^{2}}}$ and ${{d}_{\Lambda }}\triangleq {{\left| {{\omega }_{\max }} \right|}^{2}}\underset{k\in \bar{\Lambda }}{\mathop{\max }}\,\sum\limits_{l=1}^{\upsilon }{{{\sum\limits_{m\in \Lambda \backslash \mathcal{J}}{\left| \left\langle \boldsymbol{\varphi }_{m}^{(l)},\mathbf{Q}_{\mathcal{J}}^{(l)}\boldsymbol{\varphi }_{m}^{(l)} \right\rangle  \right|}}^{2}}}$. For sufficiently small noise levels, i.e., $\eta <({{c}_{\Lambda }}-{{d}_{\Lambda }})\sqrt{\frac{2}{\pi }}\upsilon $, the probability of (14) not being satisfied, denoted as $P_{mis}^{(1)}$, is bounded as 
\begin{equation}
P_{mis}^{(1)}\le \left| \mathcal{U}_{a}^{(1)} \right|{{2}^{\left| \Lambda  \right|}}\exp (-\upsilon {}\frac{\gamma _{\Lambda }^{2}}{\pi }),   
\end{equation}
where ${{\gamma }_{\Lambda }}=\frac{{{c}_{\Lambda }}-{{d}_{\Lambda }}-\eta {{\left( \sqrt{\frac{2}{\pi }\upsilon } \right)}^{-1}}}{{{c}_{\Lambda }}+{{d}_{\Lambda }}}$ and $\left| \mathcal{U}_{a}^{(1)} \right|$ is cardinality of the hypothesis set $\mathcal{U}_{a}^{(1)}$ [18].

 The parameters ${{d}_{\Lambda }}$ and ${{c}_{\Lambda }}$ can be considered as guidelines to design the spreading sequences under consideration. In particular, it is desired to design them so that ${{c}_{\Lambda }}$ is maximized while minimizing ${{d}_{\Lambda }}$ for any arbitrary support set $\Lambda $. However, the nature of this design problem is combinatorically complex. Therefore, it may not be feasible to design the sequences in a handcrafted manner. Recently, however, a neural-network (NN)-based design approach is introduced by exploiting the NN's ability to approximate complex optimization problems [32]. At this end, the above parameters can be used in the NN-based sequences design for GF-RA. 

 A complete proof of (15) with a slight change of notations can be found in Theorem 7 of [18]. Moreover, it is to be noted that the misdetection error rate is computed for a Gaussian-distributed signal in [18]. However, as the data symbols considered in this paper are taken from a sub-Gaussian distribution, (15) still serves as an upper bound.  It should be noted from (15) that the probability of signature misdetection, $P_{mis}^{(1)}$, in a single measurement instance of a symbol group decreases exponentially with $\upsilon $. 

Moreover, we can extend the results in (15), which applies to a single-symbol group measurement instance, to ${{N}_{g}}$-symbol groups of the received frame model discussed in Section III. We argue that the MMV algorithm converges to the correct support if it does not fail in more than half of ${{N}_{g}}$ group detection. This is ensured as the weight of each atom is updated by the IORLS algorithm such that atoms that have already been correctly identified are encouraged in the forthcoming wSOMP operations. With each wSOMP operations, furthermore, the weights of the incorrect atoms decrease if those atoms were not already detected in more than ${{N}_{g}}/2$ groups, discouraging their detection in the forthcoming wSOMP operations. Let ${{P}_{mis}}$ and $P_{mis}^{({{N}_{g}}/2)}$ denote the probability of misdetecting the signatures at the end of the IORLS algorithm and the probability of failure in more than ${{N}_{g}}/2$ group symbol measurements, respectively. Then, the condition for convergence of IORLS to correctly detect all atoms is given as ${{P}_{mis}}\le P_{mis}^{({{N}_{g}}/2)}$.  As proven in Appendix A, therefore, the misdetection probability ${{P}_{mis}}$ is bounded as
\begin{equation}
{{P}_{mis}}<{{K}_{3}}\exp (-({{K}_{2}}\upsilon -1){{N}_{g}}/2),
\end{equation}
where ${{K}_{2}}$ and  ${{K}_{3}}$ are constants. The result in (16) shows that the upper bound for the misdetection probability decreases exponentially with the number of symbols per group, $\upsilon $, and the number of symbol groups in a frame, ${{N}_{g}}$. This result can also be summarized as an exponentially decreasing function with the number of symbols in the GF frame as ${{N}_{c}}=\upsilon {{N}_{g}}$. Note that the failure probability upper-bounded in (16) is the upper bound on the probability of failing to detect at least one of transmitted signatures correctly (signature misdetection).

While analyzing the constraint on the property of the measurement matrices for correct support detection, we note that the second LHS term in (14) can be rewritten as
\begin{equation}
\underset{k\in \bar{\Lambda }}{\mathop{\max }}\,\sum\limits_{l=1}^{\upsilon }{{{\sum\limits_{m\in \Lambda \backslash \mathcal{J}}{\left| \left\langle \boldsymbol{\varphi }_{m}^{(l)},\mathbf{Q}_{\mathcal{J}}^{(l)}\boldsymbol{\varphi }_{k}^{(l)} \right\rangle  \right|}}^{2}}\le }\sum\limits_{l=1}^{\upsilon }{\frac{\mu _{2}^{(l)}(\mathcal{J})}{1-{{\delta }^{(l)}}(\mathcal{J})}},
\end{equation}
where  $\mu _{2}^{(l)}(\mathcal{J})$and ${{\delta }^{(l)}}(\mathcal{J})$ are the 2-Babel mutual coherence and isometry constant functions, respectively, defined on the support $\mathcal{J}$ and the matrix ${{\mathbf{\Phi }}^{(l)}}$, respectively. Similarly for the first term, we have
\begin{equation}
\underset{m\in \Lambda \backslash \mathcal{J}}{\mathop{\min }}\,\sum\limits_{l=1}^{\upsilon }{\left| \left\langle \boldsymbol{\varphi }_{m}^{(l)},\mathbf{Q}_{\mathcal{J}}^{(l)}\boldsymbol{\varphi }_{m}^{(l)} \right\rangle  \right|}\ge \upsilon -\sum\limits_{l=1}^{\upsilon }{\frac{{{\left( \mu _{2}^{(l)}(\mathcal{J}) \right)}^{2}}}{1-{{\delta }^{(l)}}(\mathcal{J})}\text{.                      }}
\end{equation}
Detailed derivation of (17) and (18) are provided in Part A and Part B of Appendices, respectively. In order to correctly detect all the supports (transmitted signatures), the constraint defined in (14) should be fulfilled by the set $\mathcal{J}=\Lambda $. Using (17) and (18) while setting $\mathcal{J}=\Lambda $, the constraint is now given as the following inequality:
\begin{equation}
\upsilon -\sum\limits_{l=1}^{\upsilon }{\frac{\left( \mu _{2}^{(l)}(\Lambda ) \right)\left( 1-\mu _{2}^{(l)}(\Lambda ) \right)}{1-{{\delta }^{(l)}}(\Lambda )}\ge\eta. }   
\end{equation}
Note that the worst-case constraint for (19) is when all the 2-Babel functions values $\{\mu _{2}^{(l)}(\Lambda )\}$ are at their maximum for all measurement matrices $\{{{\mathbf{\Phi }}^{(l)}}\}$ concurrently. In such a case, the second term on the left-hand side of (19) is at the maximum. In the next section, we show that the maximum of $\mu _{2}^{(l)}(\Lambda )$ is averaged out via multiple sequences in MSRA. Therefore, (19) is fulfilled even under high utilization factor, i.e., $L=\frac{{{N}_{a}}}{M}$. 

 \section{Simulation Results}
 \vspace*{-5pt}
In this section, we evaluate various aspects of MSRA with respect to the conventional CS-based RA schemes that employ single sequence-based spreading (SSRA). Accordingly, we attempt to demonstrate the code diversity gain, with respect to AUD, achieved by MSRA while presenting the detection performance of an MMV-based CS receiver detailed in Section III.  We have considered wideband (WB) and narrowband (NB) GF-RA systems with $N_{sc}^{d}=128$ and $N_{sc}^{d}=32$ subcarriers (SC), respectively. Spreading sequences with lengths of $M=32$ are mapped and superimposed onto 32 subcarriers in the frequency domain. Elements (chips) of each sequence are generated from a normalized independent identical complex Gaussian distribution, i.e., $\mathbb{C}\mathbb{N}\left( 0,{1}/{\sqrt{M}}\; \right)$. Furthermore, active users transmit symbols taken from a QPSK alphabet. A single-tap and a three-tap Rayleigh fading channels with a flat power profile are considered for NB and WB systems, respectively. A set of the base sequences generated from i.i.d. Gaussian distribution forms a set ${{\mathsf{\mathcal{B}}}_{s}}$ which consist  of ${{N}_{s}}=1024$ sequences, from which a pool of ${{N}_{T}}=1024$ signatures, ${{\mathsf{\mathcal{P}}}_{s}}$, is formed. Elements (chips) of the spreading sequences are generated from i.i.d. Gaussian distribution. Therefore, the radio resource scaling factor is set at $K=\frac{1024}{32}=32$ for both SSRA and MSRA. Each signature with a symbol group size of $\upsilon =4,8,\cdots ,32$ sequences is constructed from multiple sequences, as described in Section III. The performance evaluation in this paper focuses on how MSRA supports more UEs for grant-free access, relating the results to the theoretical analysis presented in the previous section. Throughout the paper, the simulation results are obtained by averaging the 5,000 Monte Carlo simulation repetitions.

First, Fig. 6 demonstrates the manner in which the non-orthogonality among transmissions is kept low by MSRA at a high scaling factor (SF). The average 2-Babel mutual coherence  ${{\mu }_{2}}(\Lambda )$ in (3) is used as a measure of non-orthogonality among the spreading signatures. The value of ${{\mu }_{2}}(\Lambda )$ indicates the worst correlation of an inactive signature with the superposed signature transmission in the support set. In particular, (19) shows the manner in which ${{\mu }_{2}}(\Lambda )$ affects the perfect recovery guarantee of the SOMP-based IORLS algorithm. As ${{\mu }_{2}}(\Lambda )$ increases, the summation term in (19) increases, and the constraint for perfect recovery (inequality) is not fulfilled.  In Fig. 6, we have compared the single-sequence spreading random access (SSRA) and MSRA in terms of  ${{\mu }_{2}}(\Lambda )$ and SF while the number of active users (a size of the support set $\Lambda $) is increased. Clearly, ${{\mu }_{2}}(\Lambda )$ increases when SF and the number of active UEs increase. The use of multiple sequences in MSRA enables the worst correlation between sequences to be averaged out which is illustrated by lower $\upsilon $ in MSRA as compared to SSRA. 
   \begin{figure}[t]
\centering
\includegraphics[width=3in,height=2.1in]{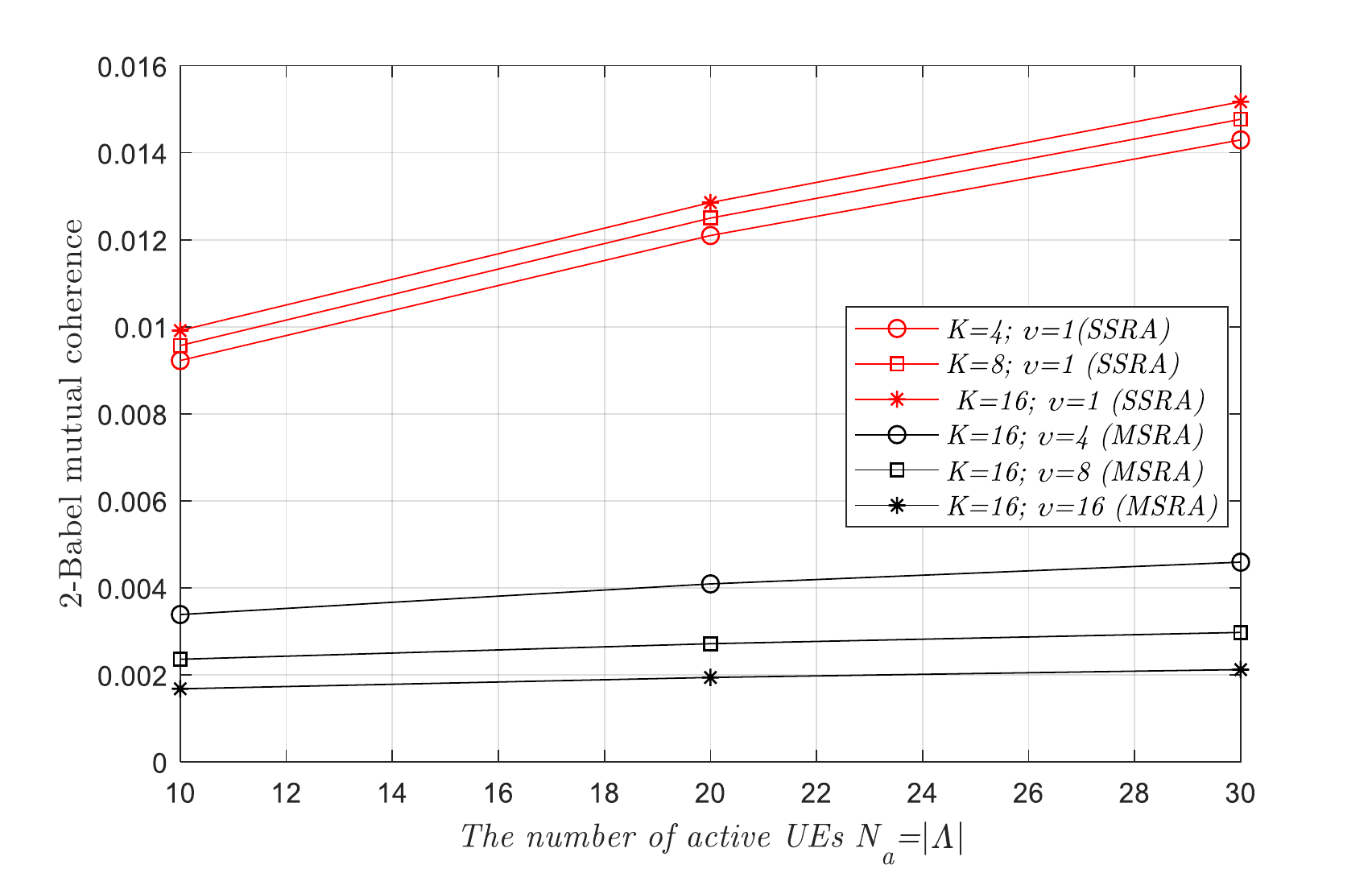}
\captionsetup{justification=centering}
\vspace*{-10pt}
\caption{\small{ 2-Babel mutual coherence value ${{\mu }_{2}}(\Lambda )$ as varying the number of active UEs, ${{N}_{a}}=\left| \Lambda  \right|$}}
\vspace*{-15pt}
\label{fig_sim}
\end{figure}

Fig. 7 illustrates the performance of MSRA in an NB system with 32 subcarriers and 3.75KHz SC spacing, implying that a 120KHz band for GF data transmission. Note that this system can be supported by the current NB-IoT standard [26]. Furthermore, it can be observed that for a carrier frequency of 1.8GHz and for device that moves at a speed of 10km/hr, the channel coherence time spans 42 OFDM symbols. In this paper, however, we consider a GF access slot with one and 32 OFDM symbols for preamble and data transmission, respectively. For this NB system, as a single-tap channel is considered the first sequence of a signature is used as a preamble sequence as discussed in Section II. Fig. 7(a) illustrates the code diversity achieved by MSRA in terms of reducing the signature misdetection rate with respect to the number of superposed active UEs. It is compared to the conventional SSRA, which is considered as a special case of MSRA with $\upsilon =1$, employing single sequence spreading [9-16] [23, 24, 27]. 

Fig. 7(a) shows the performance for increasing the number of sequences in signature $\upsilon $ while increasing the utilization factor $L={}^{{{N}_{a}}}/{}_{M}$. Note that the number of symbol groups ${{N}_{g}}={}^{{{N}_{c}}}/{}_{\upsilon }$ varies with $\upsilon $ since ${{N}_{c}}$ is fixed to 32 implying both SSRA and MSRA are applied to the same GF-RA setup. For all the cases, the CS-detection problem is cast in the MMV setup while employing the IORLS algorithm discussed in the previous section. As the number of active UEs increases, the sparsity of the measured signal decrease and hence, the misdetection rate increases with the number of active UEs in all cases. As suggested by the upper bound in (15), the detection error rate decreases exponentially with the group size. It is demonstrated in Fig. 7(a) by decrease in ${{P}_{mis}}$ with a number of sequences per signature ($\upsilon $) when $L=0.75$, i.e., there are 24 active UEs. A very low signature misdetection rate would be indicative of oracle detection, in which the receiver knows which signatures are present in the received signal and hence, the MUD problem would be transformed to an ordinary detection problem. This is shown in Fig 8(b). For $\upsilon =16$ and $\upsilon =32$, in fact, the symbol error rate achieves the oracle performance in all utilization regime except $L\ge 0.85$.  

\begin{figure}[t]
\centering
\includegraphics[width=6in,height=2.5in]{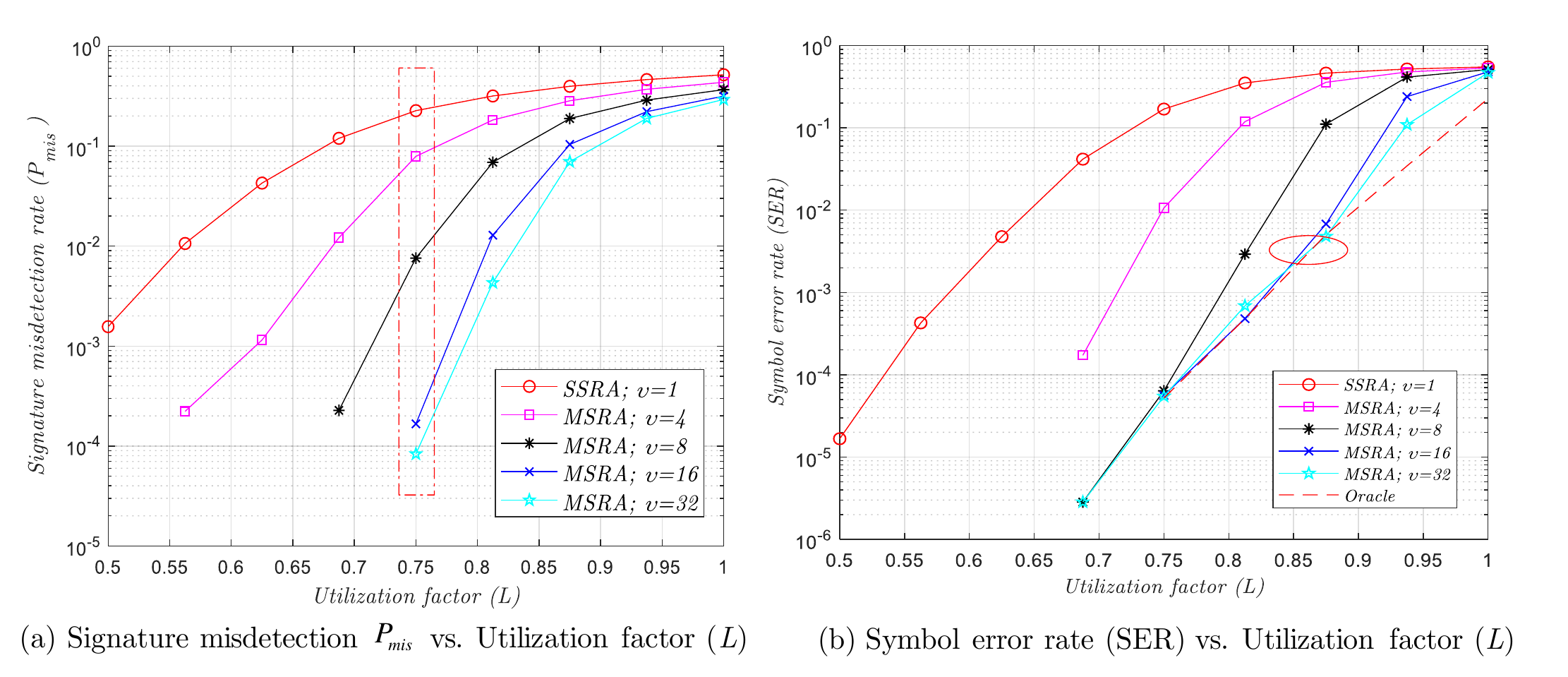}
\vspace*{-10pt}
\captionsetup{font=small,justification=centering}
\vspace*{-5pt}
\caption{\small{Performance for narrowband system: $M$=32, $N_T  $=1024, and $K$=32}}
\vspace*{-20pt}
\label{fig_sim}
\end{figure}

\begin{figure}[t]
\centering
\includegraphics[width=6in,height=2.5in]{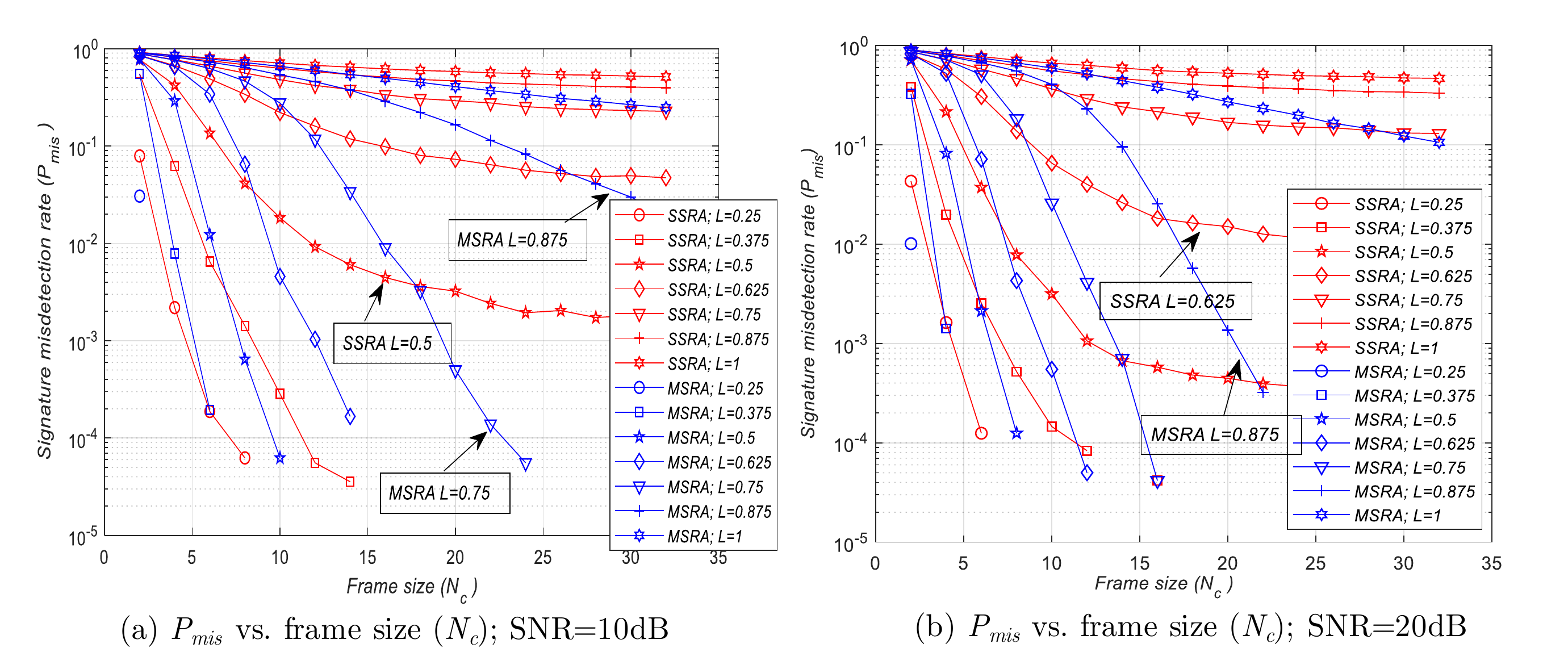}
\vspace*{-10pt}
\captionsetup{font=small,justification=centering}
\vspace*{-5pt}
\caption{\small{Performance for narrowband system with respect to frame size $N_c$: $M$=32, $N_T$ =1024, and $K$=32}}
\vspace*{-15pt}
\label{fig_sim}
\end{figure} 

In Fig. 8, the misdetection rate ${{P}_{mis}}$ is plotted against the frame size (${{N}_{c}}$) for the NB system discussed in Fig. 6. The number of multiple sequence for MSRA is set as the same as the frame size, i.e., $\upsilon ={{N}_{c}}$. We first observe that, as it is correctly predicted by equation (16), ${{P}_{mis}}$ for both MSRA and SSRA decreases exponentially with ${{N}_{c}}$ until a certain limit in utilization factor $L$ is reached.  The maximum limit in utilization factor $L$ to achieve an exponential decrease in ${{P}_{mis}}$ for MSRA and SSRA, however, is different while it also depends on the SNR level. This is a direct consequence of the constraint in (19).  For example, it can be observed from Fig. 8(a) that for SSRA when $L=0.5$ , ${{P}_{mis}}$ is no longer exponentially decreasing with ${{N}_{c}}$ since it starts to level off after ${{N}_{c}}=20$. For MSRA, however, even at $L=0.875$, ${{P}_{mis}}$ decreases exponentially while ${{N}_{c}}$ increases even if it is decreasing at slower rate than for low $L$ values. It is to be noted here that the worst Babel mutual coherence which increases with $L=\frac{{{N}_{a}}}{M}$ (as shown in Fig. 6) is much smaller for MSRA than SSRA. Therefore, the constraint in (19) is fulfilled at high utilization factor $L$ for MSRA than SSRA. Moreover, as it can be deducted from the constraint in (19), when the noise power $\eta $ decreases (i.e., SNR increases), the maximum utilization factor $L$ to achieve an exponentially decreasing ${{P}_{mis}}$  increases (compare the values of $L$ indicated by arrows in Fig. 8(a) and Fig. 8(b)). 
\begin{figure}[t]
\centering
\includegraphics[width=6in,height=2.3in]{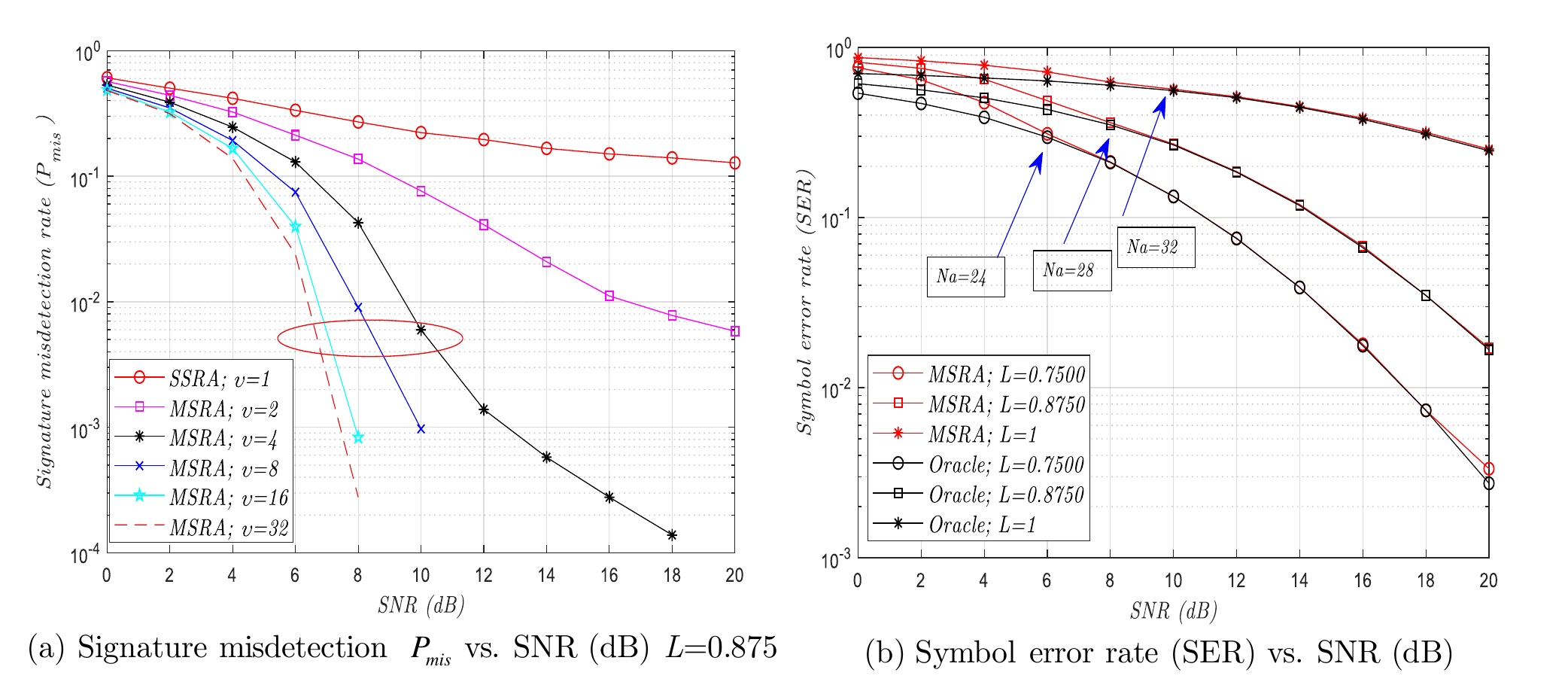}
\vspace*{-10pt}
\captionsetup{font=small, justification=centering}
\vspace*{-5pt}
\caption{\small{Performance for wideband system: $M$=32, $N_T$  =1024, and $K$=32}}
\vspace*{-15pt}
\label{fig_sim}
\end{figure} 

Fig. 9 illustrates the performance of MSRA in a WB system with $N_{SC}^{d}=128$ SCs and a 15KHz  SC spacing, implying that a 1.92MHz GF band is considered. We also considered the ${{N}_{ZC}}=127$ length ZC sequences that fits to 128 SCs by copying the first element of the ZC sequences at the last subcarrier. It is to be noted that a ZC sequences with length ${{N}_{ZC}}=331$ are employed in [27] which incurs around 3x more overhead than the proposed scheme. The ${{N}_{p}}=1024$ ZC preamble sequences are generated following the non-orthogonal preamble sequence generation scheme in [29, 30] and one-to-one associated to $N_T=1024$ sequence sets. Furthermore, as $M=32$, the radio resource scaling factor is set at $K={}^{{{N}_{T}}}/{}_{M}=32$. Moreover, in both SSRA and MSRA, active users transmit ${{N}_{c}}=128$ symbols over 32 OFDM symbols implying an OFDM symbol holds $N_g=4$ spread symbols from each user. Fig. 9 (a) illustrates the preamble misdetection rate as varying the signal-to-noise ratio (SNR) for $L=0.875$, i.e., ${{N}_{a}}=24$ active users. In the figure, it is shown that ${{P}_{mis}}$ decreases significantly with $\upsilon $ as it is predicted by (14).  In fact, SSRA performs terribly at such high utilization level $L=0.875$. Moreover, as it is predicted by (14) and (19), for  $\upsilon =4:32$ once a certain SNR level is attained, ${{P}_{mis}}$ decreases exponentially (linearly in logarithmic scale). Fig. 9 (b) provides the symbol-error rate (SER) vs SNR performance of MSRA in the same WB system at high UF $L=0.75:1$ and $\upsilon =16$. In the figure, it is shown that oracle SER is achieved by MSRA at different SNR levels for the different UF. Note that ${{\mu }_{2}}(\Lambda )$ in (19) increases with the number of active users ${{N}_{a}}=\left| \Lambda  \right|$. It confirms the prediction by (19) that the minimum SNR level required to achieve the oracle performance varies with the number of active users, i.e., 6dB, 8dB and 12dB for ${{N}_{a}}=24$, ${{N}_{a}}=28$ and ${{N}_{a}}=32$, respectively. 

 \begin{figure}[!t]
\centering
\includegraphics[width=3in,height=2.4in]{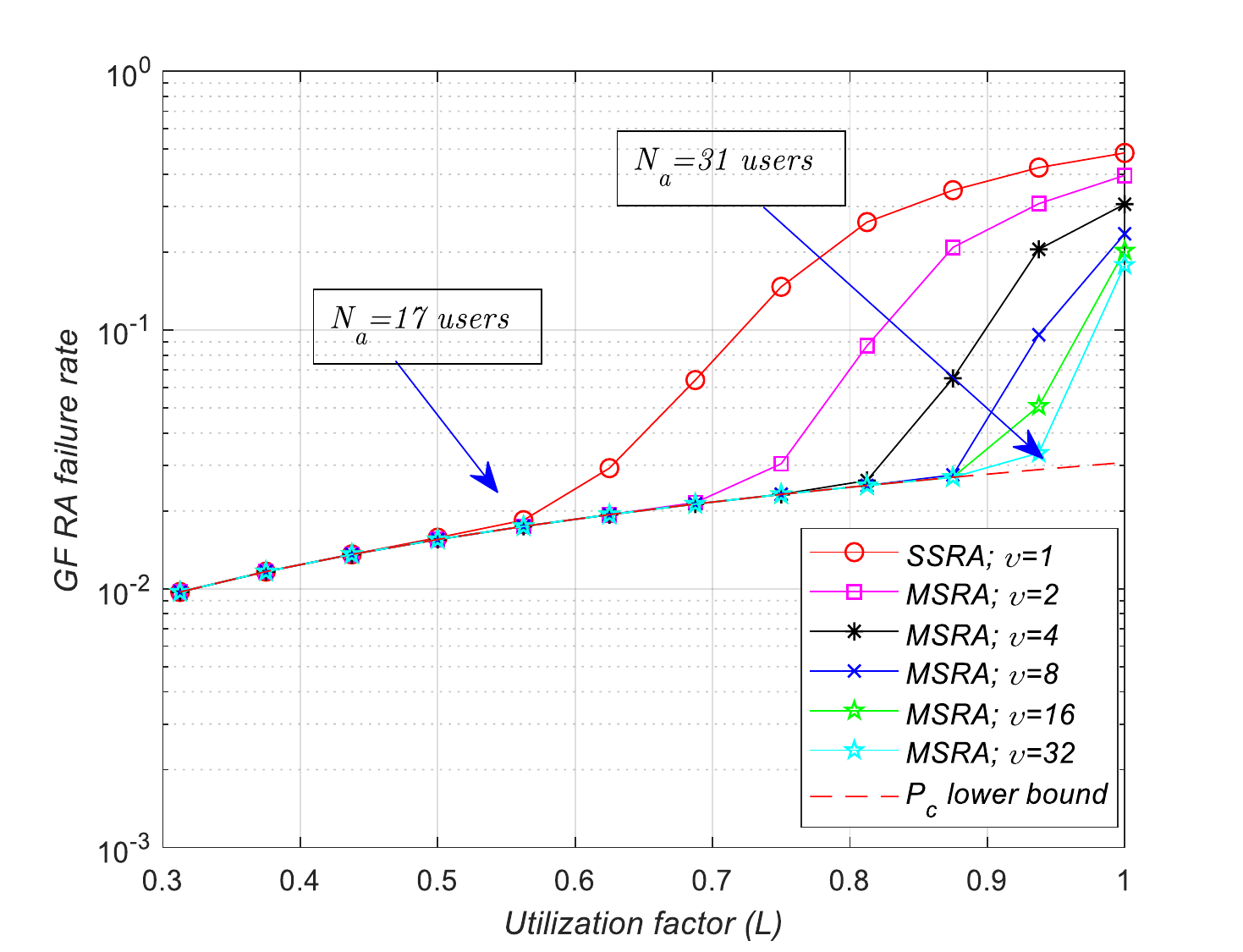}
\vspace*{-10pt}
\captionsetup{font=small, justification=centering}
\vspace*{-5pt}
\caption{\small{Failure rate vs. utilization factor (UF) in wideband MSRA: $M=32$, ${{N}_{T}}={{N}_{p}}=1024$, and SNR=10dB}}
\vspace*{-15pt}
\label{fig_sim}
\end{figure} 
 
 Finally, we evaluate the performance of MSRA in a GF-RA scenario where collision scenario is also considered. For this evaluation, we have considered the WB GF-RA configuration discussed for Fig. 9. Furthermore, our random access (GF-RA) is considered as successful if there is no collision for the user's transmission while detecting the associated preamble and data signature, i.e., user transmission identified. Otherwise, the GF-RA attempt is considered as failed. Fig. 10 presents a comparison of the GF-RA failure rates for SSRA and MSRA with varying the number of sequences in a signature. It plots the GF-RA failure rate against the utilization factor $L$. For a reference, the collision rate ${{P}_{c}}$  is also plotted since it can be considered as a lower-bound for the GF-RA failure rate that can be achieved. It can be observed that SSRA achieves this lower bound for UF below $L=0.5313$, i.e., ${{N}_{a}}=17$ users while MSRA with $\upsilon =32$ achieves this lower bound until $L=0.9688$, i.e., ${{N}_{a}}=31$. That is around 82\% gain. This result confirms the proposition in [Theorem 2.4, 19] that a well-conditioned MMV-problem with $K$-sparse vectors sharing the same support and with a sufficient number of measurements ensure a perfect support recovery (activity detection) up to $K=M-1$. Note that $M=32$ in the current simulation. From this result, we may also conjecture that the inherent code diversity in MSRA stems from using the multiple sequences which transforms the MMV-problem to be a well-conditioned MMV-problem so that the theoretical sparsity limit in [Theorem 2.4, 19] can be achieved. We want to note here that for application that requires a higher GF-RA success rate, the lower bound set by the collision rate in Fig. 10 might need to be improved. In this regard, a use of multiple preamble sequence by a user may provide diversity in RA collision from the user's perspective. It is of our future research interest to investigate the application of MSRA for ultra-reliable and low latency communication (URLLC).  
 \vspace*{-10pt}
\section{Conclusion}
\vspace*{-5pt}
The design of MSRA was presented as a NOMA scheme to enable GF-RA. Upon observing that random access collisions and active user misdetections are crucial factors in GF-RA performance, we showed that the inherent properties of MSRA simultaneously address these two factors. In particular, we showed that the use of a set of multiple sequences (signature) to spread the different symbols of a user provides code diversity. This in turn enables the MUD problem to be modeled as a well-conditioned MMV-based compressive sensing (CS) problem. The code-diversity MSRA provides is parameterized by the 2-Babel mutual coherence among the signatures. It is shown that the maximum of 2-Babel mutual coherence among signatures decreases while the number of multiple sequences in a signature in increased, which results in averaging out the MAI. Our analysis showed that the users activity misdetection decreases exponentially with the number of spreading sequences per signature. This ensures an oracle detection performance wherein the signatures (users) activity midsection is reduced to a practically negligible level, the MUD problem is effectively reduced to a simple detection problem. Simulation results showed that in a GF scenario, the proposed scheme can support up to 82\% more transmissions (users) than the conventional scheme (SSRA). The implementation possibilities of MSRA in both wideband (LTE/NR) and narrowband (NB-IoT)-based systems were discussed considering flat and frequency-selective channel estimation, respectively. In the future, we plan to investigate the application of MSRA with techniques that facilitate collision diversity to meet the requirements of ultrareliable communication.
\vspace*{-20pt}
\section{Appendix}
\vspace*{-10pt}
\subsection{Proof of (16)}  
\vspace*{-10pt}
The probability of failing in more than ${}^{{{N}_{g}}}/{}_{2}$ symbol groups is given as
\vspace*{-10pt}
\begin{align}
& p_{mis}^{({{N}_{g}}/2)}=\sum\limits_{n={}^{{{N}_{g}}}/{}_{2}}^{{{N}_{g}}}{\left( \begin{matrix}
   {{N}_{g}}  \\
   n  \\
\end{matrix} \right){{\left\{ P_{mis}^{(1)} \right\}}^{n}}{{\left\{ 1-P_{mis}^{(1)} \right\}}^{{{N}_{g}}-n}}}\text{  } \\
& <\frac{{{N}_{g}}}{2}{{\left( 2e \right)}^{{}^{{{N}_{g}}}/{}_{2}}}{{\left\{ P_{mis}^{(1)} \right\}}^{{}^{{{N}_{g}}}/{}_{2}}}{{\left\{ 1-P_{mis}^{(1)} \right\}}^{{}^{{{N}_{g}}}/{}_{2}}}\text{  }\\
& <\frac{{{N}_{g}}}{2}{{\left( 2e \right)}^{{}^{{{N}_{g}}}/{}_{2}}}{{K}_{1}}\exp (-{{K}_{2}}\upsilon {{N}_{g}}/2)(1-{{K}_{1}}\exp (-{{K}_{2}}\upsilon {{N}_{g}}/2))\text{ }    \\
&<{{K}_{3}}\exp (-({{K}_{2}}\upsilon -1){{N}_{g}}/2)\text{  }                                                                    
\end{align}
In the above derivation, (21) follows from (20), as the summation term on the right hand side is maximum when $n={{{N}_{g}}}/{2}\;$ and the upper bound on combinations function. Moreover, (22) follows from (15) and we collect constants as ${{K}_{1}}={{N}_{T}}{{2}^{\left| \Lambda  \right|}}$, ${{K}_{2}}={}^{\gamma _{\Lambda }^{2}}/{}_{\pi }$, and ${{K}_{3}}={}^{{{N}_{g}}}/{}_{2}{{\left( 2e \right)}^{{}^{{{N}_{g}}}/{}_{2}}}{{K}_{1}}$.             
\vspace*{-25pt}
\subsection{Proof of (17)} 
\vspace*{-10pt}
On the basis of the definition of $\ell$-2 norm, the second term of (14) can first be rewritten as (24) and then, subsequently bound as follows:  
\vspace*{-10pt}
\begin{align}
& \underset{k\in \bar{\Lambda }}{\mathop{\max }}\,\sum\limits_{l=1}^{\upsilon }{{{\sum\limits_{m\in \Lambda \backslash \mathcal{J}}{\left| \left\langle \boldsymbol{\varphi }_{m}^{(l)},\mathbf{Q}_{\mathcal{J}}^{(l)}\boldsymbol{\varphi }_{k}^{(l)} \right\rangle  \right|}}^{2}}}=\underset{k\in \bar{\Lambda }}{\mathop{\max }}\,\sum\limits_{l=1}^{\upsilon }{\left\| \mathbf{\Phi }_{\Lambda \backslash \mathcal{J}}^{(l)}\mathbf{Q}_{\mathcal{J}}^{(l)}\boldsymbol{\varphi }_{k}^{(l)} \right\|_{2}^{2}}     \\
& =\underset{k\in \bar{\Lambda }}{\mathop{\max }}\,\sum\limits_{l=1}^{\upsilon }{\left\| \mathbf{\Phi }_{\Lambda \backslash \mathcal{J}}^{(l)}(\mathbf{I}-\mathbf{p}_{\mathcal{J}}^{(l)})\boldsymbol{\varphi }_{k}^{(l)} \right\|_{2}^{2}}\text{ }\\
& \le \underset{k\in \bar{\Lambda }}{\mathop{\max }}\,\sum\limits_{l=1}^{\upsilon }{\left\| {{\left( \mathbf{\Phi }_{\Lambda \backslash \mathcal{J}}^{(l)} \right)}^{\text{H}}}\boldsymbol{\varphi }_{k}^{(l)} \right\|_{2}^{2}}+\sum\limits_{l=1}^{\upsilon }{\left\| {{\left( \mathbf{\Phi }_{\Lambda \backslash \mathcal{J}}^{(l)} \right)}^{\dagger }}\mathbf{\Phi }_{\mathcal{J}}^{(l)} \right\|_{2}^{2}}\sum\limits_{l=1}^{\upsilon }{\left\| {{(\mathbf{\Phi }_{\mathcal{J}}^{(l)})}^{H}}\boldsymbol{\varphi }_{k}^{(l)} \right\|_{2}^{2}}\\
& =\sum\limits_{l=1}^{\upsilon }{\frac{\mu _{2}^{(l)}(\mathcal{J})}{1-{{\delta }^{(l)}}(\mathcal{J})}\text{  }} 
\end{align}
In the above derivation, (25) considers $\mathbf{Q}_{\mathcal{J}}^{(l)}=\mathbf{I}-\mathbf{p}_{\mathcal{J}}^{(l)}$, where $\mathbf{p}_{\mathcal{J}}^{(l)}$ is a projection matrix. Moreover, (26) employs the definition of the projection matrix and assumes the upper bound of the 2-Babel function ${{\mu }_{2}}(\centerdot )$ and the isometry constant function $\delta (\centerdot )$, given in [18].
\vspace*{-24pt}
\subsection{Proof of (18)} 
\vspace*{-10pt}
Similar to proof of (17), the first term of (14) can be rewritten as (28) by substituting $\mathbf{Q}_{\mathcal{J}}^{(l)}$ with $\mathbf{I}-\mathbf{p}_{\mathcal{J}}^{(l)}$ and then, subsequently bounded as follows: 
\begin{align}
& \underset{m\in \Lambda \backslash \mathcal{J}}{\mathop{\min }}\,\sum\limits_{l=1}^{\upsilon }{\left| \left\langle \boldsymbol{\varphi }_{m}^{(l)},\mathbf{Q}_{\mathcal{J}}^{(l)}\boldsymbol{\varphi }_{m}^{(l)} \right\rangle  \right|}=\underset{m\in \Lambda \backslash \mathcal{J}}{\mathop{\min }}\,\sum\limits_{l=1}^{v}{\left| \left\langle \mathbf{I}-\mathbf{P}_{\mathcal{J}}^{(l)}\boldsymbol{\varphi }_{m}^{(l)},\boldsymbol{\varphi }_{m}^{(l)} \right\rangle  \right|}\text{        }\\
& =\underset{m\in \Lambda \backslash \mathcal{J}}{\mathop{\min }}\,\sum\limits_{l=1}^{\upsilon }{\left| 1-{{(\boldsymbol{\varphi }_{m}^{(l)})}^{\text{H}}}\mathbf{\Phi }_{\mathcal{J}}^{(l)}{{\left( {{\left( \mathbf{\Phi }_{\mathcal{J}}^{(l)} \right)}^{\text{H}}}\mathbf{\Phi }_{\mathcal{J}}^{(l)} \right)}^{-1}}{{\left( \boldsymbol{\Phi }_{\mathcal{J}}^{(l)} \right)}^{\text{H}}}\boldsymbol{\varphi }_{m}^{(l)} \right|}\text{   }
\end{align}
\begin{align}
&\ge \upsilon -\sum\limits_{l=1}^{\upsilon }{\frac{{{\left( \mu _{2}^{(l)}(\mathcal{J}) \right)}^{2}}}{1-{{\delta }^{(l)}}(\mathcal{J})}\text{  }} 
\end{align}



%


\ifCLASSOPTIONcaptionsoff
  \newpage
\fi

\end{document}